\documentclass{JFM-FLM_Au}

\lefttitle{T. Wu et al.}
\righttitle{Exact formulas for velocity-gradient moments in isotropic turbulence}

\title{Exact formulas for arbitrary order velocity-gradient moments in isotropic turbulence}

\author{Tong Wu\aff{1}, Chensheng Luo\aff{2, 3}, Le Fang\aff{3}, Michael Wilczek\aff{1}}

\affiliation{\aff{1}Theoretical Physics I, University of Bayreuth, 95440 Bayreuth,
Germany
\aff{2}Research Institute of Aero-Engine, Beihang University, Beijing 100191, China
\aff{3}Sino-French Carbon Neutrality Research Center, Ecole Centrale de Pekin, Beihang Univetrsity, Beijing 100191, China}

\corresau{Tong.Wu@uni-bayreuth.de, Michael.Wilczek@uni-bayreuth.de}

\begin{document}
\maketitle

\begin{abstract}
Statistical moments of velocity gradients provide fundamental information on the small-scale properties of turbulence.
In this work, we propose a systematic method to derive exact expressions for statistical moments of arbitrary order for both longitudinal and transverse velocity gradients in isotropic turbulence. The approach is applicable to both compressible and incompressible flows and expresses the moments in terms of invariants of the velocity gradient tensor.
The derivation combines isotropic tensor theory, orientational averaging, and an algorithmic implementation, enabling the computation of high-order moments in a unified framework. We show that longitudinal velocity gradient moments of order higher than three depend not only on $\mathrm{tr}(\mathsfbi{S}^2)$, which is proportional to the dissipation rate, but also on $\mathrm{tr}(\mathsfbi{S}^3)$, which reflects strain self-amplification, where $\mathsfbi{S}$ denotes the strain-rate tensor.
The resulting theoretical expressions are validated through comparisons with existing theoretical results and direct numerical simulations.
\end{abstract}


\section{Introduction}
The statistics of velocity gradients, which encode information about small-scale turbulent structures, play a crucial role in turbulence research \citep{Pope,wallace2009twenty,meneveau2011lagrangian,johnson2024multiscale}. In particular, high-order moments of velocity gradients are widely used to characterize small-scale intermittency \citep[e.g.]{Ishihara2007,Ishihara2009}. In isotropic turbulence, the statistical moments of velocity gradients must be invariant under rotations and reflections. As a result, these moments take the form of isotropic tensors, whose structure can be expressed in terms of a finite set of scalar invariants, providing a systematic framework for analysing the statistical properties of small-scale turbulence \citep{Taylor1935, robertson1940invariant, BookBatchelor}. 

Let $A_{ij} = \partial u_i / \partial x_j$ denote the velocity gradient tensor, and let us, as an example, consider second-order statistics.
Under the assumption of statistical isotropy, the second-order moment 
$\langle A_{ij} A_{kl} \rangle$ must take the form of a fourth-rank isotropic tensor. 
It can therefore be expressed as
\begin{equation}
\langle A_{ij} A_{kl} \rangle 
= \alpha \, \delta_{ij} \delta_{kl} 
+ \beta \, \delta_{ik} \delta_{jl} 
+ \gamma \, \delta_{il} \delta_{jk},
\end{equation}
where $\delta_{ij}$ is the second-order identity tensor, and 
$\alpha$, $\beta$ and $\gamma$ are scalar coefficients \citep{robertson1940invariant, weyl1946classical, BookBatchelor}. 
In homogeneous, incompressible, isotropic turbulence, the coefficients can be expressed as $\alpha = \gamma = - \frac{1}{4}\beta = -\frac{\langle S_{ij}S_{ij} \rangle}{15}$ where $S_{ij} = \tfrac{1}{2}(A_{ij} + A_{ji})$ is the strain-rate tensor. Contraction of this tensor yields the well-known expressions for the second
order moments of longitudinal and transverse velocity-gradient \citep{Pope},
\begin{equation}
\langle A_{11}^2 \rangle = \frac{\varepsilon}{15\nu}, 
\qquad 
\langle A_{21}^2 \rangle = \frac{2\varepsilon}{15\nu},
\end{equation}
where $\varepsilon = 2\nu S_{ij} S_{ij}$ is the kinetic energy dissipation rate and
$\nu$ is the kinematic viscosity. 
The quantity $S_{ij} S_{ij}= \mathrm{tr}(\mathsfbi{S}^2)$ is a rotationally invariant scalar.

For higher-order moments, the construction of isotropic tensor representations become increasingly complex due to the rapid growth in the number of independent contraction terms. For an $n$th-order moment, this number is $(2n-1)!!$ \citep{phan1994isotropic}. While the number of independent contractions increases from $3$ to $15$ for third-order moments \citep{phan1994isotropic, Pope}, the corresponding invariant representations remain manageable. 
Indeed, the invariant expression for the third-order longitudinal velocity-gradient moment was derived by \cite{Champagne1978} (in the appendix of the paper). 
At fourth order, however, the number of independent contractions rises sharply to $105$, resulting in a large linear system that must be solved to determine the coefficients associated with the scalar invariants. 
Direct derivations therefore become prohibitively laborious, as shown by \cite{hierro2003fourth}.

Several approaches have been proposed to avoid this difficulty. 
\cite{betchov1956inequality} expressed $A_{11}$ in the strain principal-axis frame using strain eigenvalues and orientation angles, followed by orientation averaging to obtain longitudinal velocity-gradient moments up to fourth order. Recently, \cite{yang2026relation} extended this framework to arbitrary orders of longitudinal velocity-gradient moments. However, this approach does not naturally generalize to transverse gradients.
\cite{siggia1981invariants} proposed an alternative method based on Gaussian ensembles, which also allows a direct derivation of the fourth-order moment of $A_{11}$. 
However, this approach relies on the proportionality between $\langle A_{11}^{2n} \rangle$ and $\langle [\mathrm{tr}(\mathsfbi{S}^2)]^n \rangle$, which holds only for $n=1$ and $2$, and therefore cannot be generalized to moments of order higher than four. 
For the same reason, the assumption made by \cite{boschung2015exact} that all even-order longitudinal moments $\langle A_{11}^{2n} \rangle$ depend solely on $\langle \varepsilon^n \rangle$ is not rigorously true.

In this paper, we develop an analytical framework to derive exact expressions for arbitrary-order statistical moments of both longitudinal and transverse velocity gradients in homogeneous isotropic turbulence, applicable to both compressible and incompressible flows. 
The resulting formulas are expressed in terms of scalar invariants of the velocity gradient tensor. By combining the ideas of isotropic tensor representations and orientational averaging, 
our method circumvents the need to solve large linear systems and enables the systematic derivation of longitudinal and transverse velocity-gradient moments of arbitrary order.

This paper is organized as follows. 
Section~\ref{Sec:moments_A11} presents the derivation of the $n$th-order moments of longitudinal velocity gradients. 
Section~\ref{Sec:moments_A21} extends the methodology to transverse gradients. 
The numerical simulations used to validate the analytical results are presented in Section~\ref{Sec:numerical}. 
Conclusions and perspectives are provided in Section~\ref{Sec:conclu}.

\section{Moments of longitudinal velocity gradients}\label{Sec:moments_A11}
This section aims to express the $n$th-order moments of the longitudinal velocity gradient in isotropic turbulence in terms of invariants. Due to isotropy, the statistics of longitudinal velocity gradients are invariant under rotation. Without loss of generality, we focus on $\langle A_{11}^n \rangle$ where $\langle \bullet \rangle$ is the ensemble average (equivalent to the spatial average due to the assumed isotropy and implied homogeneity). The moments of other longitudinal components are statistically equivalent.

\subsection{Invariants of the strain-rate tensor}\label{Sec:invariant_S}

Since $A_{11} = S_{11}$, the statistics of $A_{11}$ are governed solely by $\mathsfbi{S}$. In three-dimensional (3D) turbulence, the symmetric tensor $\mathsfbi{S} \in \mathbb{R}^{3 \times 3}$ has six independent components. In isotropic turbulence, statistical properties must remain invariant under any rotation of the coordinate system. This rotational symmetry effectively removes three degrees of freedom associated with arbitrary orientations, so that the statistics of $\mathsfbi{S}$ can be characterized by three independent scalar invariants. 

A convenient choice of such invariants is given by the traces of powers of $\mathsfbi{S}$, which are rotationally invariant scalars. According to the Cayley–Hamilton theorem \citep{lang1987linear}, for any $3\times 3$ matrix $\mathsfbi{B}$, we have:
\begin{equation}\label{Eq:CH_theorem}
\mathsfbi{B}^3 - (\operatorname{tr} \mathsfbi{B}) \mathsfbi{B}^2 + \frac{1}{2} \left[ (\operatorname{tr} \mathsfbi{B})^2 - \operatorname{tr}(\mathsfbi{B}^2) \right] \mathsfbi{B} - \det(\mathsfbi{B}) \mathsfbi{I} = \mathsfbi{O}.
\end{equation}
where $\mathsfbi{I}$ is the $3\times 3$ identity matrix and $\mathsfbi{O}$ is the zero matrix. The determinant of $\mathsfbi{B}$ can be expressed in terms of traces as:
\begin{equation}
\det(\mathsfbi{B}) = \frac{1}{6} \left[ (\operatorname{tr} \mathsfbi{B})^3 - 3\operatorname{tr}(\mathsfbi{B}^2)\operatorname{tr}(\mathsfbi{B}) + 2\operatorname{tr}(\mathsfbi{B}^3) \right].
\end{equation}
The identity \eqref{Eq:CH_theorem} implies that for any $n \geq 4$, the quantity $\operatorname{tr}(\mathsfbi{B}^n)$ can be written as a function of $\operatorname{tr}(\mathsfbi{B})$, $\operatorname{tr}(\mathsfbi{B}^2)$, and $\operatorname{tr}(\mathsfbi{B}^3)$.
Therefore, the three fundamental invariants are:
\[
\operatorname{tr}(\mathsfbi{S}), \quad \operatorname{tr}(\mathsfbi{S}^2), \quad \operatorname{tr}(\mathsfbi{S}^3).
\]
In incompressible turbulence, $\operatorname{tr}(\mathsfbi{S}) = 0$, so only $\operatorname{tr}(\mathsfbi{S}^2)$ and $\operatorname{tr}(\mathsfbi{S}^3)$ remain fundamental invariants.

\subsection{Key idea for expressing $\langle A_{11}^n \rangle$}

In isotropic turbulence, the $n$th-order moment of the longitudinal velocity gradient is identical in all directions. Therefore, $\langle A_{11}^n \rangle$ is equal to the average of $n$th-order moments of longitudinal velocity gradients taken over all directions. Since ensemble and orientational averaging are interchangeable, we can first average over orientations and then perform the ensemble average. The key implication of this idea is:
\begin{equation}
    \langle A_{11}^n \rangle_{\text{space}}  = \langle \langle A_{\parallel}^n \rangle_{\text{orien}} \rangle_{\text{space}},
\end{equation}
where $A_{\parallel}$ denotes the longitudinal velocity gradient along an arbitrary direction $\boldsymbol{e}$. Specifically, it is derived as
\[
A_{\parallel} := e_i S_{ij} e_j,
\]
where $\boldsymbol{e}$ is a unit vector uniformly distributed over the unit sphere. Here and in the following, we adopt the Einstein summation convention over repeated indices. Consequently, the $n$th-order moment can be expressed as:
\begin{equation}
    \langle A_{11}^n \rangle =
    \left\langle \left\langle e_{i_1}e_{i_2} \cdots e_{i_{2n}} \right\rangle_{\text{orien}} 
    S_{i_1i_2} \cdots S_{i_{2n-1}i_{2n}} \right\rangle.
\end{equation}

The computation proceeds in three steps:
\begin{enumerate}
    \item Compute the orientational average of the unit-vector product:
    \[
    \left\langle e_{i_1} e_{i_2} \cdots e_{i_{2n}} \right\rangle_{\text{orien}}.
    \]
    This yields a fully symmetric isotropic tensor of rank $2n$.
    
    \item Contract this result with the tensor product of $n$ strain-rate tensors:
    \[
    S_{i_1 i_2} \cdots S_{i_{2n-1} i_{2n}}.
    \]
    \item Take the ensemble average of the resulting scalar.
\end{enumerate}

\subsection{Derivation of the formula of $\langle A_{11}^n \rangle$}
\subsubsection{Orientational average of the unit-vector product}

We first consider the average of products of components of a unit vector:
\[
M^{(n)}_{i_1\cdots i_{2n}}
:= \big\langle e_{i_1}\cdots e_{i_{2n}}\big\rangle_{\text{orien}},
\]
where $\boldsymbol{e}$ is a unit vector uniformly distributed over the surface of the unit sphere in $d$ dimensions. This average yields a fully rotationally invariant tensor of rank $2n$.

It is a classical result \citep{phan1994isotropic, eyink2006multi} that:
\begin{equation}\label{Eq:average_e_prodcut}
M^{(n)}_{i_1\cdots i_{2n}}= C_n^d \sum_{\text{all pairings}} \;\prod_{\text{paired indices} \, \{a,b\}} \delta_{i_a i_b},
\end{equation}
where 
\begin{equation} \label{Eq:Cnd_general}
C_n^d = \frac{1}{d(d+2)(d+4)\cdots (d+2n-2)}
\end{equation}
with $d$ the spatial dimension.
In 3D turbulence (\( d = 3 \)), the isotropic factor $C_n^d$ simplifies to a double factorial:
\begin{equation} \label{Eq:Cnd_3D}
C_n^3 = \frac{1}{(2n + 1)!!}.
\end{equation}
The summation in Eq. \eqref{Eq:average_e_prodcut} runs over all distinct pairings (i.e., partitions into $n$ disjoint
pairs) of the index set $\{1,2,\ldots,2n\}$, and each pairing contributes a product
of Kronecker deltas $\delta_{i_a i_b}$ over all paired indices $(a,b)$.
For example, for $n = 2$, there are $3$ pairings: $\delta_{i_1 i_2} \delta_{i_3 i_4}$, $\delta_{i_1 i_3} \delta_{i_2 i_4}$, and $\delta_{i_1 i_4} \delta_{i_2 i_3}$, yielding
\begin{align}
M^{(2)}_{i_1 i_2 i_3 i_4} &= \frac{1}{15}\Big(\delta_{i_1 i_2} \delta_{i_3 i_4}
+ \delta_{i_1 i_3} \delta_{i_2 i_4}
+ \delta_{i_1 i_4} \delta_{i_2 i_3}\Big).
\end{align}
For $n=3$, there are 15 pairings, and
\begin{equation}
\begin{aligned}\label{Eq:average_e_prodcut_n3}
M^{(3)}_{i_1 i_2 i_3 i_4 i_5 i_6} = \frac{1}{105} \Big(&
\delta_{i_1 i_2} \delta_{i_3 i_4} \delta_{i_5 i_6} +
\delta_{i_1 i_2} \delta_{i_3 i_5} \delta_{i_4 i_6} +
\delta_{i_1 i_2} \delta_{i_3 i_6} \delta_{i_4 i_5} \\
&+
\delta_{i_1 i_3} \delta_{i_2 i_4} \delta_{i_5 i_6} +
\delta_{i_1 i_3} \delta_{i_2 i_5} \delta_{i_4 i_6} +
\delta_{i_1 i_3} \delta_{i_2 i_6} \delta_{i_4 i_5} \\
&+
\delta_{i_1 i_4} \delta_{i_2 i_3} \delta_{i_5 i_6} +
\delta_{i_1 i_4} \delta_{i_2 i_5} \delta_{i_3 i_6} +
\delta_{i_1 i_4} \delta_{i_2 i_6} \delta_{i_3 i_5} \\
&+
\delta_{i_1 i_5} \delta_{i_2 i_3} \delta_{i_4 i_6} +
\delta_{i_1 i_5} \delta_{i_2 i_4} \delta_{i_3 i_6} +
\delta_{i_1 i_5} \delta_{i_2 i_6} \delta_{i_3 i_4} \\
&+
\delta_{i_1 i_6} \delta_{i_2 i_3} \delta_{i_4 i_5} +
\delta_{i_1 i_6} \delta_{i_2 i_4} \delta_{i_3 i_5} +
\delta_{i_1 i_6} \delta_{i_2 i_5} \delta_{i_3 i_4}
\Big).
\end{aligned}
\end{equation}

\subsubsection{Contraction with $\mathsfbi{S}$ product}

We now evaluate the contraction:
\begin{equation}\label{Eq:contraction}
    \left\langle e_{i_1} e_{i_2} \cdots e_{i_{2n}} \right\rangle_{\text{orien}} \, S_{i_1 i_2} \cdots S_{i_{2n-1} i_{2n}}.
\end{equation}
Substituting Eq.~\eqref{Eq:average_e_prodcut} into Eq.~\eqref{Eq:contraction}, we obtain a sum of contractions involving products of $S_{ij}$, where each term reduces to a product of traces of powers of $\mathsfbi{S}$.

To illustrate the contraction procedure, consider the case $n=3$:
\begin{equation}\label{Eq:contraction_n3}
\begin{aligned}
    M^{(3)}_{i_1 i_2 i_3 i_4 i_5 i_6} S_{i_1 i_2} S_{i_3 i_4} S_{i_5 i_6} &= \frac{1}{105} \Big(
\delta_{i_1 i_2} \delta_{i_3 i_4} \delta_{i_5 i_6} +
\delta_{i_1 i_2} \delta_{i_3 i_5} \delta_{i_4 i_6} +
\delta_{i_1 i_2} \delta_{i_3 i_6} \delta_{i_4 i_5} \\
&+
\delta_{i_1 i_3} \delta_{i_2 i_4} \delta_{i_5 i_6} +
\delta_{i_1 i_3} \delta_{i_2 i_5} \delta_{i_4 i_6} +
\delta_{i_1 i_3} \delta_{i_2 i_6} \delta_{i_4 i_5} \\
&+
\delta_{i_1 i_4} \delta_{i_2 i_3} \delta_{i_5 i_6} +
\delta_{i_1 i_4} \delta_{i_2 i_5} \delta_{i_3 i_6} +
\delta_{i_1 i_4} \delta_{i_2 i_6} \delta_{i_3 i_5} \\
&+
\delta_{i_1 i_5} \delta_{i_2 i_3} \delta_{i_4 i_6} +
\delta_{i_1 i_5} \delta_{i_2 i_4} \delta_{i_3 i_6} +
\delta_{i_1 i_5} \delta_{i_2 i_6} \delta_{i_3 i_4} \\
&+
\delta_{i_1 i_6} \delta_{i_2 i_3} \delta_{i_4 i_5} +
\delta_{i_1 i_6} \delta_{i_2 i_4} \delta_{i_3 i_5} +
\delta_{i_1 i_6} \delta_{i_2 i_5} \delta_{i_3 i_4}
\Big)S_{i_1 i_2} S_{i_3 i_4} S_{i_5 i_6}
    \end{aligned}
\end{equation}

The 15 pairings lead to three distinct types of contractions:
\begin{itemize}
    \item All three $\mathsfbi{S}$ tensors are contracted independently, e.g., 
    \[
    \delta_{i_1 i_2} \delta_{i_3 i_4} \delta_{i_5 i_6} S_{i_1 i_2} S_{i_3 i_4} S_{i_5 i_6} = \mathrm{tr}(\mathsfbi{S})^3.
    \]
    There is only 1 such term.
    
    \item One $\mathsfbi{S}$ contracts with itself, and the remaining two are cross-contracted, e.g.,
    \[
    \delta_{i_1 i_2} \delta_{i_3 i_5} \delta_{i_4 i_6} S_{i_1 i_2} S_{i_3 i_4} S_{i_5 i_6} = \mathrm{tr}(\mathsfbi{S}) \, \mathrm{tr}(\mathsfbi{S}^2).
    \]
    This type contributes 6 terms.

    \item All three $\mathsfbi{S}$ tensors are fully contracted together, e.g.,
    \[
    \delta_{i_1 i_3} \delta_{i_2 i_5} \delta_{i_4 i_6} S_{i_1 i_2} S_{i_3 i_4} S_{i_5 i_6} = \mathrm{tr}(\mathsfbi{S}^3).
    \]
    This type contributes 8 terms.
\end{itemize}
Therefore, the total contraction is given by:
\begin{equation}\label{Eq:contraction_A11}
    M^{(3)}_{i_1 \cdots i_6} S_{i_1 i_2} S_{i_3 i_4} S_{i_5 i_6} 
    = \frac{1}{105} \left( \mathrm{tr}(\mathsfbi{S})^3 + 6\,\mathrm{tr}(\mathsfbi{S})\,\mathrm{tr}(\mathsfbi{S}^2) + 8\,\mathrm{tr}(\mathsfbi{S}^3) \right).
\end{equation}
These three types of contractions are naturally classified by the integer partitions
of $3$, which specify how the three $\mathsfbi{S}$ tensors ($S_{i_1 i_2}$, $S_{i_3 i_4}$, and $S_{i_5 i_6} $) are grouped under index
contraction. Each partition corresponds to a distinct trace structure, while the
associated coefficient is determined by the number of admissible Kronecker-delta
pairings consistent with that grouping.
\begin{itemize}
\item $3 = 1 + 1 + 1$ partitions the three $\mathsfbi{S}$ tensors into three
subsets, each containing a single tensor. There is only one such partition,
$\{\{S_{i_1 i_2}\}, \{S_{i_3 i_4}\}, \{S_{i_5 i_6}\}\}$, which corresponds to the case
in which all three tensors are contracted independently and yields the invariant
$\mathrm{tr}(\mathsfbi{S})^3$. Moreover, only a single Kronecker-delta pairing,
$\delta_{i_1 i_2}\delta_{i_3 i_4}\delta_{i_5 i_6}$, is admissible with this contraction
pattern. As a result, the coefficient of $\mathrm{tr}(\mathsfbi{S})^3$ in
Eq.~\eqref{Eq:contraction_A11} is simply $C_3^3$.
\item $3 = 1 + 2$ partitions the three $\mathsfbi{S}$ tensors into two
subsets, one containing a single tensor and the other containing two tensors.
There are three such partitions,
$\{\{S_{i_1 i_2}\}, \{S_{i_3 i_4}, S_{i_5 i_6}\}\}$,
$\{\{S_{i_3 i_4}\}, \{S_{i_1 i_2}, S_{i_5 i_6}\}\}$, and
$\{\{S_{i_5 i_6}\}, \{S_{i_1 i_2}, S_{i_3 i_4}\}\}$.
Each of these partitions corresponds to a contraction in which one tensor is
contracted independently while the remaining two are contracted together,
yielding the invariant $\mathrm{tr}(\mathsfbi{S})\,\mathrm{tr}(\mathsfbi{S}^2)$.
For each partition, there are two Kronecker-delta pairings compatible
with this contraction pattern. Consequently, the coefficient of
$\mathrm{tr}(\mathsfbi{S})\,\mathrm{tr}(\mathsfbi{S}^2)$ in
Eq.~\eqref{Eq:contraction_A11} is $3 \times 2 \times C_3^3 = 6C_3^3$.
\item $3 = 3$ corresponds to the partition in which all three
$\mathsfbi{S}$ tensors belong to a single subset,
$\{\{S_{i_1 i_2}, S_{i_3 i_4}, S_{i_5 i_6}\}\}$.
This partition corresponds to the contraction yielding the invariant $\mathrm{tr}(\mathsfbi{S}^3)$.
In this case, eight distinct Kronecker-delta pairings are compatible with this contraction pattern. As a result, the coefficient of
$\mathrm{tr}(\mathsfbi{S}^3)$ is
$1 \times 8 \times C_3^3 = 8C_3^3$.
\end{itemize}

In the general case of $n$ $\mathsfbi{S}$ tensors, the integer partitions of $n$ can be systematically encoded using the Bell polynomials \citep{bell1934exponential, comtet2012advanced}.
Let \(B_{n}(x_1,x_2,\dots, x_n)\) be the complete Bell polynomials:
\begin{equation}\label{eq:Bell-def}
B_{n}(x_1,x_2,\dots, x_n)
=\sum_{\substack{m_1,\dots,m_n\ge 0\\ \sum_{k=1}^n k m_k=n}}
\frac{n!}{\prod_{k=1}^n m_k!} \prod_{k=1}^n \Big(\frac{x_k}{k!}\Big)^{m_k},
\end{equation}
where $x_k$ corresponds to the presence of a subset with $k$ elements in a given partition, and $m_k$ denotes the number of such subsets, satisfying $\sum_{k=1}^n k m_k = n$. There are $\frac{n!}{\prod_{k=1}^n m_k!} \prod_{k=1}^n \Big(\frac{1}{k!}\Big)^{m_k}$ distinct ways to partition a set of $n$ elements in this $m_k$-structure.

In our context, a subset with $k$ elements corresponds to a contraction involving $k$ $\mathsfbi{S}$ tensors, resulting in a term of the form $\operatorname{tr}(\mathsfbi{S}^k)$. From basic combinatorial arguments, we know that each subset with $k$ $\mathsfbi{S}$ tensors corresponds to \(2^{k-1}(k-1)!\) pairings of Kronecker-deltas. We absorb this multiplicity into the Bell polynomial inputs by setting
\[
x_k \;=\; 2^{k-1}(k-1)!\,\operatorname{tr}(\mathsfbi{S}^k).
\]  
Using the identity \(\sum_{k=1}^n (k-1)m_k = n - \sum_{k=1}^n m_k\), we rewrite the product as
\begin{equation}\label{Eq:Bell-def_trace}
\prod_{k=1}^n \Big(\frac{x_k}{k!}\Big)^{m_k}
= \frac{2^{\,n-\sum_{k=1}^n m_k}}{\prod_{k=1}^n k^{m_k}}
\prod_{k=1}^n [\operatorname{tr}(\mathsfbi{S}^k)]^{m_k}.
\end{equation}
Substituting expression~\eqref{Eq:Bell-def_trace} into~\eqref{eq:Bell-def} and multiplying by the isotropic factor $C_n^3$ (Eq. \eqref{Eq:Cnd_3D}) gives
\begin{equation}\label{Eq:A_11_n_orien}
\begin{aligned}
\langle A_{||}^n\rangle_{\text{orien}} &= M^{(n)}_{i_1\cdots i_{2n}} S_{i_1 i_2} \cdots S_{i_{2n-1} i_{2n}} \\
& = \frac{1}{(2n+1)!!} \sum_{\substack{m_1,\dots,m_n\ge 0\\ \sum k m_k=n}}
2^{\,n-\sum_{k=1}^{n} m_k}\,
\frac{n!}{\prod_{k=1}^n k^{\,m_k}\,m_k!}\;
\prod_{k=1}^{n}\big[\operatorname{tr}(S^k)\big]^{m_k}.
\end{aligned}
\end{equation}

\subsubsection{Ensemble average}

The final expression for $\langle A_{11}^n \rangle$ is obtained by taking the ensemble average of Eq.~\eqref{Eq:A_11_n_orien}:
\begin{equation}\label{Eq:A_11_n}
\langle A_{11}^n\rangle
= \frac{1}{(2n+1)!!} \sum_{\substack{m_1,\dots,m_n\ge 0\\ \sum k m_k=n}}
2^{\,n-\sum m_k}\,
\frac{n!}{\prod_{k=1}^n k^{\,m_k}\,m_k!}\;
\left\langle \prod_{k=1}^{n}  \big[\operatorname{tr}(S^k)\big]^{m_k} \right\rangle.
\end{equation}

As noted in Section~\ref{Sec:invariant_S}, the invariants \(\operatorname{tr}(\mathsfbi{S}^k)\) for \(k > 3\) can be written as polynomial combinations of \(\operatorname{tr}(\mathsfbi{S})\), \(\operatorname{tr}(\mathsfbi{S}^2)\), and \(\operatorname{tr}(\mathsfbi{S}^3)\). Moreover, for incompressible turbulence, the condition \(\operatorname{tr}(\mathsfbi{S}) = 0\) further simplifies these expressions. These simplifications are carried out using a symbolic Python code, which is available in the accompanying \href{https://cocalc.com/share/public_paths/78ee59efe8b5ac10ea4d7f68c2d337e7413afea1}{JFM notebook}.

\subsection{Analytical expressions of $\langle A_{11}^n \rangle$}
In this section, we present analytical expressions of $\langle A_{11}^n \rangle$ for selected values of $n$. While it is impractical to list all orders explicitly, our general formula~\eqref{Eq:A_11_n} and accompanying symbolic implementation (available in the \href{https://cocalc.com/share/public_paths/78ee59efe8b5ac10ea4d7f68c2d337e7413afea1}{JFM notebook}) allow the computation of any desired moment.

\subsubsection{Incompressible turbulence}

In this subsection, we present the expressions computed from our symbolic code for $\langle A_{11}^n\rangle$ with $n$ ranging from 2 to 10 in incompressible turbulence:
\begin{align}
\langle A_{11}^2\rangle &= \frac{2}{15} \langle\operatorname{tr}(\mathsfbi{S}^2)\rangle, \label{Eq:exact_formular_A11_incompressible_2} \\
\langle A_{11}^3\rangle &= \frac{8}{105} \langle\operatorname{tr}(\mathsfbi{S}^3)\rangle, \\
\langle A_{11}^4\rangle &= \frac{4}{105} \langle\operatorname{tr}(\mathsfbi{S}^2)^2\rangle, \\
\langle A_{11}^5\rangle &= \frac{32}{693} \langle\operatorname{tr}(\mathsfbi{S}^2)\operatorname{tr}(\mathsfbi{S}^3)\rangle, \\
\langle A_{11}^6\rangle &= \frac{40}{3003} \langle\operatorname{tr}(\mathsfbi{S}^2)^3\rangle + \frac{128}{9009} \langle\operatorname{tr}(\mathsfbi{S}^3)^2\rangle, \\
\langle A_{11}^7\rangle &= \frac{32}{1287} \langle\operatorname{tr}(\mathsfbi{S}^2)^2\operatorname{tr}(\mathsfbi{S}^3)\rangle, \\
\langle A_{11}^8\rangle &=  \frac{112}{21879}\langle\operatorname{tr}(\mathsfbi{S}^2)^4\rangle + \frac{1024}{65637}\langle\operatorname{tr}(\mathsfbi{S}^2)\operatorname{tr}(\mathsfbi{S}^3)^2\rangle, \\
\langle A_{11}^9\rangle &= \frac{1792}{138567} \langle\operatorname{tr}(\mathsfbi{S}^2)^3\operatorname{tr}(\mathsfbi{S}^3)\rangle + \frac{4096}{1247103}\langle\operatorname{tr}(\mathsfbi{S}^3)^3\rangle, \\
\langle A_{11}^{10}\rangle &= \frac{96}{46189}\langle\operatorname{tr}(\mathsfbi{S}^2)^5\rangle + \frac{5120}{415701} \langle\operatorname{tr}(\mathsfbi{S}^2)^2\operatorname{tr}(\mathsfbi{S}^3)^2\rangle 
\end{align}

The expressions for the second and third-order moments agree with well-known results \citep{Pope}. The fourth-order moment is consistent with previously published findings \citep{betchov1956inequality, siggia1981invariants, hierro2003fourth}, and the sixth-, eighth-, and tenth-order moments agree with the recent results reported by \cite{yang2026relation}. We argue that the even-order expressions reported in \cite{boschung2015exact}
are valid only up to $n<6$, as their derivation relies on the implicit assumption
that even-order moments of the longitudinal velocity gradient depend solely on
the statistics of $\operatorname{tr}(\mathsfbi{S}^2)$. Our results show that
$\operatorname{tr}(\mathsfbi{S}^3)$ contributes for all $n \ge 6$, indicating that
this assumption is no longer sufficient at higher orders.

These results further imply that all moments of the longitudinal velocity gradient are completely determined by the joint probability density function (PDF) of $\operatorname{tr}(\mathsfbi{S}^2)$ and $\operatorname{tr}(\mathsfbi{S}^3)$ for incompressible turbulence.

\subsubsection{Compressible turbulence}
In this subsection, we present the expressions computed from our symbolic code for $\langle A_{11}^n\rangle$ with $n$ ranging from 2 to 8 in compressible turbulence:
\begin{align}
\langle A_{11}^2\rangle &=\frac{1}{15} \langle\operatorname{tr}(\mathsfbi{S})^2\rangle+ \frac{2}{15} \langle\operatorname{tr}(\mathsfbi{S}^2)\rangle,\label{Eq:exact_formular_A11_compressible_2}  \\
\langle A_{11}^3\rangle &= \frac{1}{105} \langle \mathrm{tr}(\mathsfbi{S})^3 \rangle + \frac{2}{35}\, \langle\mathrm{tr}(\mathsfbi{S}) \,\mathrm{tr}(\mathsfbi{S}^2)\rangle + \frac{8}{105}\, \langle \mathrm{tr}(\mathsfbi{S}^3) \rangle, \label{Eq:exact_formular_A11_compressible_3} \\
\langle A_{11}^4\rangle &= \frac{1}{105} \langle \mathrm{tr}(\mathsfbi{S})^4 \rangle -\frac{4}{105}\langle \mathrm{tr}(\mathsfbi{S})^2 \mathrm{tr}(\mathsfbi{S}^2)\rangle +\frac{4}{105} \langle \mathrm{tr}(\mathsfbi{S}^2)^2\rangle +
\frac{32}{315}\langle \mathrm{tr}(\mathsfbi{S}) \mathrm{tr}(\mathsfbi{S}^3)\rangle, \\
\langle A_{11}^5\rangle &=\frac{1}{99}\left\langle \mathrm{tr}(\mathsfbi{S})^5 \right\rangle
- \frac{4}{77}\left\langle \mathrm{tr}(\mathsfbi{S})^3\,\mathrm{tr}(\mathsfbi{S}^2) \right\rangle
+ \frac{4}{231}\left\langle \mathrm{tr}(\mathsfbi{S})\,\mathrm{tr}(\mathsfbi{S}^2)^2 \right\rangle
+ \frac{16}{231}\left\langle \mathrm{tr}(\mathsfbi{S})^2\,\mathrm{tr}(\mathsfbi{S}^3) \right\rangle  \nonumber \\
\qquad &\quad + \frac{32}{693}\left\langle \mathrm{tr}(\mathsfbi{S}^2)\,\mathrm{tr}(\mathsfbi{S}^3) \right\rangle, \\
\langle A_{11}^6\rangle &= \frac{5}{819}\left\langle \mathrm{tr}(\mathsfbi{S})^6 \right\rangle
-\frac{74}{3003}\left\langle \mathrm{tr}(\mathsfbi{S})^4 \,\mathrm{tr}(\mathsfbi{S}^2) \right\rangle
-\frac{4}{143}\left\langle \mathrm{tr}(\mathsfbi{S})^2 \,\mathrm{tr}(\mathsfbi{S}^2)^2 \right\rangle
+ \frac{40}{3003}\left\langle \mathrm{tr}(\mathsfbi{S}^2)^3 \right\rangle \nonumber \\
\qquad &\quad + \frac{32}{1001}\left\langle \mathrm{tr}(\mathsfbi{S})^3 \,\mathrm{tr}(\mathsfbi{S}^3) \right\rangle
+ \frac{64}{1001}\left\langle \mathrm{tr}(\mathsfbi{S})\,\mathrm{tr}(\mathsfbi{S}^2)\,\mathrm{tr}(\mathsfbi{S}^3) \right\rangle
+ \frac{128}{9009}\left\langle \mathrm{tr}(\mathsfbi{S}^3)^2 \right\rangle,\\
\langle A_{11}^7\rangle &= 
\frac{49}{19305}\left\langle \mathrm{tr}(\mathsfbi{S})^7 \right\rangle
- \frac{34}{6435}\left\langle \mathrm{tr}(\mathsfbi{S})^5 \, \mathrm{tr}(\mathsfbi{S}^2) \right\rangle + \frac{56}{3861}\left\langle \mathrm{tr}(\mathsfbi{S})^4 \, \mathrm{tr}(\mathsfbi{S}^3) \right\rangle \nonumber \\ 
\qquad &\quad - \frac{4}{117}\left\langle \mathrm{tr}(\mathsfbi{S})^3 \, \mathrm{tr}(\mathsfbi{S}^2)^2 \right\rangle + \frac{32}{1287}\left\langle \mathrm{tr}(\mathsfbi{S})^2 \, \mathrm{tr}(\mathsfbi{S}^2)\, \mathrm{tr}(\mathsfbi{S}^3) \right\rangle  + \frac{8}{1287}\left\langle \mathrm{tr}(\mathsfbi{S}) \, \mathrm{tr}(\mathsfbi{S}^2)^3 \right\rangle \nonumber \\ 
\qquad &\quad
+ \frac{128}{3861}\left\langle \mathrm{tr}(\mathsfbi{S}) \, \mathrm{tr}(\mathsfbi{S}^3)^2 \right\rangle  + \frac{32}{1287}\left\langle \mathrm{tr}(\mathsfbi{S}^2)^2 \, \mathrm{tr}(\mathsfbi{S}^3) \right\rangle, \\
\langle A_{11}^8\rangle &= \frac{329}{328185}\left\langle \mathrm{tr}(\mathsfbi{S})^8 \right\rangle
- \frac{56}{29835}\left\langle \mathrm{tr}(\mathsfbi{S})^6 \, \mathrm{tr}(\mathsfbi{S}^2) \right\rangle  + \frac{3392}{328185}\left\langle \mathrm{tr}(\mathsfbi{S})^5 \, \mathrm{tr}(\mathsfbi{S}^3) \right\rangle \nonumber \\ 
\qquad &\quad - \frac{24}{2431}\left\langle \mathrm{tr}(\mathsfbi{S})^4 \, \mathrm{tr}(\mathsfbi{S}^2)^2 \right\rangle  - \frac{1280}{65637}\left\langle \mathrm{tr}(\mathsfbi{S})^3 \, \mathrm{tr}(\mathsfbi{S}^2)\, \mathrm{tr}(\mathsfbi{S}^3) \right\rangle  - \frac{32}{1989}\left\langle \mathrm{tr}(\mathsfbi{S})^2 \, \mathrm{tr}(\mathsfbi{S}^2)^3 \right\rangle \nonumber \\ 
\qquad &\quad + \frac{2560}{65637}\left\langle \mathrm{tr}(\mathsfbi{S})^2 \, \mathrm{tr}(\mathsfbi{S}^3)^2 \right\rangle + \frac{256}{7293}\left\langle \mathrm{tr}(\mathsfbi{S}) \, \mathrm{tr}(\mathsfbi{S}^2)^2 \, \mathrm{tr}(\mathsfbi{S}^3) \right\rangle + \frac{112}{21879}\left\langle \mathrm{tr}(\mathsfbi{S}^2)^4 \right\rangle \nonumber \\ 
\qquad &\quad + \frac{1024}{65637}\left\langle \mathrm{tr}(\mathsfbi{S}^2)\, \mathrm{tr}(\mathsfbi{S}^3)^2 \right\rangle 
\end{align}

The expressions for the second- and third-order moments (Eq. \eqref{Eq:exact_formular_A11_compressible_2} and \eqref{Eq:exact_formular_A11_compressible_3}) are consistent with the results reported by \cite{yang2022low}. As expected, the moments of the compressible case reduce to the ones of the incompressible case when setting $\mathrm{tr}(\mathsfbi{S})=0$.

Furthermore, these results show that, in compressible turbulence, all moments of the longitudinal velocity gradient are fully determined by the joint PDF of $\operatorname{tr}(\mathsfbi{S})$, $\operatorname{tr}(\mathsfbi{S}^2)$, and $\operatorname{tr}(\mathsfbi{S}^3)$.

\section{Moments of transverse velocity gradients}\label{Sec:moments_A21}

This section aims to express the $n$th-order moments of transverse velocity gradients in isotropic turbulence in terms of invariants. Without loss of generality, we focus on the component $\langle A_{21}^n \rangle$, as all other transverse components are statistically equivalent.

\subsection{Invariants of the velocity gradient tensor}\label{Sec:invariant_A}
To evaluate $\langle A_{21}^n \rangle$, we study scalar invariants constructed from the velocity gradient tensor $\mathsfbi{A}$. In 3D turbulence, $\mathsfbi{A}$ has nine components, but isotropy removes three rotational degrees of freedom, thus the statistics of $\mathsfbi{A}$ can be characterized by six independent scalar invariants.
The following set of fundamental invariants is used in this work:
\begin{equation}\label{Eq:invariants_A}
\begin{aligned}
I_1 &= \operatorname{tr}(\mathsfbi{A}), I_2 =   \operatorname{tr}(\mathsfbi{A}^2), I_3 = \operatorname{tr}(\mathsfbi{A} \mathsfbi{A}^T), I_4 = \operatorname{tr}(\mathsfbi{A}^3), I_5=\operatorname{tr}(\mathsfbi{A} \mathsfbi{A} \mathsfbi{A}^T), \\
I_6 &= \operatorname{tr}(\mathsfbi{A} \mathsfbi{A} \mathsfbi{A}^T \mathsfbi{A}^T).
\end{aligned}
\end{equation}
For incompressible turbulence, $I_1 = 0$, reducing the number of fundamental invariants to five. 

Furthermore, in turbulence research, statistical isotropy is typically formulated under the assumption of statistical homogeneity, which imposes constraints on ensemble-averaged velocity-gradient invariants. For incompressible
turbulence, \cite{betchov1956inequality} showed that these constraints read
\begin{align}
\big\langle I_2 \big\rangle &= 0, \label{Eq:Betchov_incompressible}\\
\big\langle I_4 \big\rangle &= 0.
\end{align}
For compressible turbulence, the corresponding homogeneity constraints were derived in recent studies~\citep{yang2022low, carbone2022only} and take the form
\begin{align}
\big\langle I_1 \big\rangle = 0, \\
\big\langle I_2 - I_1^2 \big\rangle = 0, \label{Eq:Betchov_compressible} \\
\Big\langle 
    I_4 
    - \frac{3}{2}\,I_1 I_2
    + \frac{1}{2}\,I_1^3
\Big\rangle = 0.
\end{align}
It was further shown in~\cite{carbone2022only} that these exhaust all homogeneity constraints. As a consequence, Betchov constraints need to be taken into account only when
evaluating the second-order longitudinal gradient moment (i.e.,
$\langle A_{21}^2 \rangle$). Higher-order constraints beyond third order do not exist, while the first- and third-order transverse gradient moments vanish identically.

Throughout this work, we employ a set of invariants based on the velocity
gradient tensor $\mathsfbi{A}$. An equivalent and widely used formulation instead
expresses the invariants in terms of the strain-rate tensor $\mathsfbi{S}$ and the
rotation-rate tensor $\mathsfbi{W}$~\citep{pope1975more, pennisi1987irreducibility, carbone2022only},
defined as
\begin{equation}\label{Eq:invariants_WS}
        I_1' = \operatorname{tr}(\mathsfbi{S}), I_2' = \operatorname{tr}(\mathsfbi{S}^2), I_3' = \operatorname{tr}(\mathsfbi{W}^2), I_4' = \operatorname{tr}(\mathsfbi{S}^3), I_5' = \operatorname{tr}(\mathsfbi{S} \mathsfbi{W}^2), I_6' =  \operatorname{tr}(\mathsfbi{S}^2 \mathsfbi{W}^2).
\end{equation}
These two sets of invariants are fully equivalent and can be transformed into one
another. The explicit relationships between the two sets, as well as the expressions
for velocity-gradient moments in terms of the invariants
\eqref{Eq:invariants_WS}, are provided in Appendix~\ref{Sec:Invariants_S_W}.

\subsection{Derivation of the formula of $\langle A_{21}^n \rangle$}\label{Sec:A21_derivation}

Following the approach in Sec.~\ref{Sec:moments_A11}, the $n$th-order moment 
$\langle A_{21}^n\rangle$ is computed by first averaging the transverse velocity-gradient 
moments over all orientations, and then taking the spatial average:
\begin{equation}
    \langle A_{21}^n \rangle_{\text{space}} = \langle \langle A_{\perp}^n \rangle_{\text{orien}} \rangle_{\text{space}}.
\end{equation}
where $A_{\perp}$ denotes the transverse velocity gradient in an arbitrary direction. Specifically, let $\boldsymbol{u}$ be a unit vector uniformly distributed over the unit sphere, and let $\boldsymbol{v}$ be a unit vector uniformly distributed over the unit circle in the plane perpendicular to $\boldsymbol{u}$. Then the transverse velocity gradient is given by
\begin{equation*}
    A_{\perp} = u_i A_{ij} v_j.
\end{equation*}
In turbulent flows, isotropy includes invariance under reflections, which implies that all odd-order transverse moments vanish. Therefore, we restrict our attention to even-order moments, i.e., $\langle A_{21}^{2n} \rangle$. These can be expressed as:
\begin{equation}\label{Eq:A21}
    \langle A_{21}^{2n} \rangle_{\text{space}} = \langle \langle u_{i_1}v_{j_1} ...u_{i_{2n}} v_{j_{2n}} \rangle_{\text{orien}} \, A_{i_1j_1}...A_{i_{2n}j_{2n}}\rangle_{\text{space}}.
\end{equation}

\subsubsection{Orientational average and contraction with products of $\mathsfbi{A}$}\label{Sec:contraction_A21_AAT}
We begin with the orientational average of mixed products of the two unit vectors:
\begin{equation}\label{Eq:A21_Tn}
    T^{(n)}_{i_1 j_1 \cdots i_{2n} j_{2n}} := 
\left\langle u_{i_1} v_{j_1} \cdots u_{i_{2n}} v_{j_{2n}} \right\rangle_{\text{orien}}.
\end{equation}
This average can be computed by first averaging over all directions of $\boldsymbol{v}$ with fixed $\boldsymbol{u}$, and then averaging over all directions of $\boldsymbol{u}$:
\begin{equation}\label{Eq:average_uv_products}
\left\langle u_{i_1} v_{j_1} \cdots u_{i_{2n}} v_{j_{2n}} \right\rangle_{\text{orien}} 
= \left\langle u_{i_1} \cdots u_{i_{2n}} 
\left\langle v_{j_1} \cdots v_{j_{2n}} \mid \boldsymbol{u} \right\rangle_{\text{orien, $\boldsymbol{v}$}} 
\right\rangle_{\text{orien, $\boldsymbol{u}$}}.
\end{equation}
Since $\boldsymbol{v}$ is uniformly distributed on the unit circle orthogonal to $\boldsymbol{u}$, 
its moment expansion follows the same structure as Eq.~\eqref{Eq:average_e_prodcut}, 
which gives the average of products of unit vector components in $d$-dimensional space. 
In our case, although the ambient space is three-dimensional, 
$\boldsymbol{v}$ lies in the two-dimensional subspace perpendicular to $\boldsymbol{u}$. 
Therefore, we set $d=2$ in the formula, and the Kronecker delta 
$\delta_{i_a i_b}$ used in Eq.~\eqref{Eq:average_e_prodcut}—which represents 
the identity tensor in the $d$-dimensional space—must be replaced by 
the projection tensor $P_{i_a i_b} = \delta_{i_a i_b} - u_{i_a} u_{i_b}$, 
which serves as the identity operator on the 2D plane orthogonal to $\boldsymbol{u}$ 
within the 3D space. Consequently, the classical result in Eq.~\eqref{Eq:average_e_prodcut} applies with $d=2$ and 
$\delta_{ij}$ replaced by $P_{ij}$, yielding:
\begin{equation}\label{Eq:orien_average_trans_v}
    \left\langle v_{j_1} \cdots v_{j_{2n}} \mid \boldsymbol{u} \right\rangle_{\text{orien, $\boldsymbol{v}$}} 
    = C_n^2 \sum_{\text{all pairings}} \prod_{\text{paired indices} \, \{a,b\}} P_{j_a j_b}
\end{equation}
with $C_n^2 = \frac{1}{2^n n!}$. 

Substituting Eq.~\eqref{Eq:orien_average_trans_v} into Eq.~\eqref{Eq:average_uv_products} and \eqref{Eq:A21_Tn}, we obtain
\begin{equation}\label{Eq:A21_implement_step1}
A_{i_1j_1}...A_{i_{2n}j_{2n}} T^{(n)}_{i_1 j_1 \cdots i_{2n} j_{2n}} = C_n^2 \left\langle \sum_{\text{all pairings}} \prod_{\text{paired indices} \, \{a,b\}}  \omega_{j_a} P_{j_a j_b} \omega_{j_b}\right\rangle_{\text{orien, $\boldsymbol{u}$}} 
\end{equation}
with $\omega_j = A_{ij} u_i$. Noting that $\omega_{j_a} P_{j_a j_b} \omega_{j_b} = \boldsymbol{\omega}^T \mathsfbi{P} \boldsymbol{\omega}$, each pairing contributes an identical scalar factor. Since the number of distinct pairings of $2n$ indices is
$(2n-1)!!$, the summation \eqref{Eq:A21_implement_step1} reduces to 
\begin{equation}\label{Eq:A21_implement_step1_simplify}
A_{i_1j_1}...A_{i_{2n}j_{2n}} T^{(n)}_{i_1 j_1 \cdots i_{2n} j_{2n}} = C_n^2 (2n-1)!! \left\langle (\boldsymbol{\omega}^T \mathsfbi{P} \boldsymbol{\omega})^n\right\rangle_{\text{orien, $\boldsymbol{u}$}} 
\end{equation}
The orientational average $\left\langle (\boldsymbol{\omega}^T \mathsfbi{P} \boldsymbol{\omega})^n\right\rangle_{\text{orien, $\boldsymbol{u}$}}$ can be expressed as a linear combination of isotropic tensors of the form
\(
\langle u_{i_1} \cdots u_{i_{2m}} \rangle_{\text{orien}}
\) where $m$ is an integer satisfying $m \le n$. Applying Eq.~\eqref{Eq:average_e_prodcut} and performing the
contraction then yields
Eq.~\eqref{Eq:A21_implement_step1_simplify}
in terms of traces involving products of
$\mathsfbi{A}$ and $\mathsfbi{A}^T$.

As an explicit example, we compute the orientational average for \( n = 1 \). Setting $n=1$ in 
Eq.~\eqref{Eq:A21_implement_step1_simplify}, we obtain
\begin{equation}\label{Eq:A21_2}
    A_{i_1j_1}A_{i_{2}j_{2}} T^{(1)}_{i_1 j_1 i_{2} j_{2}}=  \frac{1}{2} \left\langle \boldsymbol{\omega}^T \mathsfbi{P} \boldsymbol{\omega}\right\rangle_{\text{orien}} 
\end{equation}
We now evaluate the orientational average:
\begin{equation}
\begin{aligned}
     \left\langle \boldsymbol{\omega}^T \mathsfbi{P} \boldsymbol{\omega}\right\rangle_{\text{orien}} &= \left\langle A_{i_1j_1}u_{i_1}(\delta_{j_1j_2}-u_{j_1}u_{j_2})A_{i_2j_2}u_{i_2} \right\rangle_{\text{orien}}\\
     &= A_{i_1j_1}A_{i_2j_2} \langle u_{i_1}u_{i_2} \rangle_{\text{orien}}\delta_{j_1j_2} - A_{i_1j_1}A_{i_2j_2}\langle u_{i_1} u_{i_2} u_{j_1} u_{j_2}\rangle_{\text{orien}} \\
     &= A_{i_1j_1}A_{i_2j_2} \delta_{j_1j_2} M^{(1)}_{i_1i_2} - A_{i_1j_1}A_{i_2j_2}M^{(2)}_{i_1i_2j_1j_2} \\
     &= \frac{1}{3} A_{i_1j_1}A_{i_2j_2} \delta_{j_1j_2} \delta_{i_1i_2} - \frac{1}{15}A_{i_1j_1}A_{i_2j_2}(\delta_{i_1 i_2} \delta_{j_1j_2}
+ \delta_{i_1 j_1} \delta_{i_2 j_2}
+ \delta_{i_1 j_2} \delta_{i_2 j_1}) \\
     &= \frac{4}{15} \operatorname{tr}(\mathsfbi{A} \mathsfbi{A}^T)-\frac{1}{15}(\operatorname{tr}(\mathsfbi{A}))^2 - \frac{1}{15}\operatorname{tr}(\mathsfbi{A}^2).
\end{aligned}
\end{equation}
Substituting this result into Eq.~\eqref{Eq:A21_2}, we finally arrive at
\begin{equation}
    A_{i_1j_1}A_{i_{2}j_{2}} T^{(1)}_{i_1 j_1 i_{2} j_{2}} = \frac{2}{15} \operatorname{tr}(\mathsfbi{A} \mathsfbi{A}^T)  -\frac{1}{30}(\operatorname{tr}(\mathsfbi{A}))^2  - \frac{1}{30}\operatorname{tr}(\mathsfbi{A}^2). 
\end{equation}

\subsubsection{Rewriting in terms of $I_1,\dots,I_6$}
As shown in Section~\ref{Sec:invariant_A}, any traces involving products of $\mathsfbi{A}$ and $\mathsfbi{A}^T$ can be expressed in terms of the six fundamental invariants $I_1,\dots,I_6$ defined in~\eqref{Eq:invariants_A}.
For general $n$, the contraction ~\eqref{Eq:A21_implement_step1_simplify} contains $2n$ factors of $\mathsfbi{A}$ (or $\mathsfbi{A}^{T}$) and hence is a homogeneous scalar polynomial of degree $2n$ in the entries of $\mathsfbi{A}$.
We therefore seek a representation of the form
\begin{equation}\label{Eq:A21_I_form}
A_{i_1j_1}...A_{i_{2n}j_{2n}} T^{(n)}_{i_1 j_1 \cdots i_{2n} j_{2n}}
= \sum_{\substack{\boldsymbol{\alpha}=(\alpha_1, \alpha_2, \alpha_3, \alpha_4, \alpha_5, \alpha_6)\\||\boldsymbol{\alpha}||_w=2n}}
c_{\boldsymbol{\alpha}}
\, I_1^{\alpha_1} I_2^{\alpha_2} I_3^{\alpha_3} I_4^{\alpha_4} I_5^{\alpha_5} I_6^{\alpha_6},
\end{equation}
where the degrees of the six invariants are
\[
\deg I_1 = 1,\quad
\deg I_2 = \deg I_3 = 2,\quad
\deg I_4 = \deg I_5 = 3,\quad
\deg I_6 = 4,
\]
so that the admissible multi-indices satisfy the weighted-degree constraint
\[
||\boldsymbol{\alpha}||_w := \alpha_1 + 2\alpha_2 + 2\alpha_3 + 3\alpha_4 + 3\alpha_5 + 4\alpha_6 = 2n.
\]
In the incompressible case $\operatorname{tr}(\mathsfbi{A})=0$, we have $I_1=0$ and thus restrict to $\alpha_1=0$.

The admissible monomials span the ansatz space for the invariant expansion. To determine the coefficients $c_{\boldsymbol{\alpha}}$, we can expand both the contraction and the ansatz in terms of the nine components of $\mathsfbi{A}$ and equate the coefficients of all resulting monomials, which determines $c_{\boldsymbol{\alpha}}$.

\subsubsection{Ensemble average}
The final expression for $\langle A_{21}^n \rangle$ is obtained by taking the ensemble average of \eqref{Eq:A21_I_form}. 
As explained in Sec.~\ref{Sec:invariant_A}, the Betchov constraints (Eqs.~\eqref{Eq:Betchov_incompressible} and~\eqref{Eq:Betchov_compressible}) must be taken into account when evaluating the second-order moment $\langle A_{11}^2 \rangle$.

\subsection{Symbolic implementation}
Section~\ref{Sec:A21_derivation} gives an exact formula for
$\langle A_{21}^{n}\rangle$. A direct symbolic implementation of that derivation is, however, impractical, because the number of contractions (in Eq.~\eqref{Eq:A21_implement_step1}) grows of order $(4n-1)!!$, making the computation rapidly intractable for $n\gtrsim 5$. To determine the invariant coefficients $c_{\boldsymbol{\alpha}}$ without explicitly expanding Eq.~\eqref{Eq:A21_implement_step1}, we instead use a sampling-and-solve strategy based on random integer matrices $\mathsfbi{A}$. The procedure proceeds as follows.

\begin{itemize}
  \item \textit{Random sampling.} 
  
        Generate a random integer matrix
        $\mathsfbi{A}$.  
        For the incompressible case, impose
        $\operatorname{tr}\mathsfbi{A}=0$.
        
  \item \textit{Computation of orientational average.} 

Expanding $(\boldsymbol{\omega}^T \mathsfbi{P} \boldsymbol{\omega})^n$ yields a polynomial in the components of the unit vector $u=(u_1,u_2,u_3)$, so that its orientational average can be written as
\begin{equation}
\left\langle (\boldsymbol{\omega}^T \mathsfbi{P} \boldsymbol{\omega})^n \right\rangle_{\mathrm{orien}}
=
\sum_{a,b,c} K_{abc}
\left\langle u_1^a u_2^b u_3^c \right\rangle_{\mathrm{orien}},
\end{equation}
where the coefficients $K_{abc}$ depend on the entries of $\mathsfbi{A}$. By isotropy, any monomial moment with at least one odd exponent vanishes. Therefore,
\begin{equation}
\left\langle u_1^a u_2^b u_3^c \right\rangle_{\mathrm{orien}} = 0,
\qquad
\text{if any of } a,b,c \text{ is odd}.
\end{equation}
For even exponents, writing $a=2a'$, $b=2b'$, and $c=2c'$, the moment is obtained from the contraction formula in Eq.~\eqref{Eq:average_e_prodcut}, giving
\begin{equation}
\left\langle u_1^{2a'} u_2^{2b'} u_3^{2c'} \right\rangle_{\mathrm{orien}}
=
C_m^3(2a'-1)!!(2b'-1)!!(2c'-1)!!
\end{equation}
with $m=a'+b'+c'$. 

  \item \textit{Evaluation of invariant monomials.}  

For the same matrix $\mathsfbi{A}$, evaluate all admissible monomials $I_1^{\alpha_1}\cdots I_6^{\alpha_6}$
of weighted degree $2n$. These values form one integer row vector
$\boldsymbol{r}$.

\item \textit{Assembly and exact solution of the linear system.}

Repeating the above procedure for several random matrices produces a linear system
$\mathsfbi{R}\boldsymbol{c}=\boldsymbol{y},$
where each row of $\mathsfbi{R}$ contains the evaluated invariant monomial vector $\boldsymbol{r}$ and each entry of $\boldsymbol{y}$ is the corresponding exact orientational average. To reduce memory usage, each newly generated row is tested for linear independence against the previously accepted rows, and only independent rows are retained. The assembly stops once the number of accepted rows equals the number of admissible monomials. The resulting system is then solved, which yields the exact coefficients $c_{\boldsymbol{\alpha}}$.
        
  \item \textit{Consistency check.}  

The resulted coefficients are substituted back into the invariant expansion and tested on additional randomly generated matrices. Exact agreement for these extra samples provides a consistency check.
  
\end{itemize}

This sampling-and-solve strategy avoids the factorial growth of a direct symbolic expansion, and remains tractable at least up to $n=16$ in the compressible case and $n=20$ in the incompressible case. The full implementation is available in the
\href{https://cocalc.com/share/public_paths/78ee59efe8b5ac10ea4d7f68c2d337e7413afea1}{JFM notebook}.

\subsection{Analytical expressions of $\langle A_{21}^n \rangle$}

In this section, we present analytical expressions for $\langle A_{21}^n \rangle$ at selected values of $n$.
While it is impractical to list expressions for all orders, our symbolic implementation, available in the
\href{https://cocalc.com/share/public_paths/78ee59efe8b5ac10ea4d7f68c2d337e7413afea1}{JFM Notebook}, enables the computation of moments of arbitrary order (in practice, the maximum attainable order is constrained by computational cost).

\subsubsection{Incompressible turbulence}

In this subsection, we present the expressions computed from our symbolic code for $\langle A_{21}^n\rangle$ with $n=2,4,6$ in incompressible turbulence:
\begin{align}
\big\langle A_{21}^2 \big\rangle
    &= \frac{2}{15}\langle I_3\rangle,\\
    \big\langle A_{21}^4 \big\rangle
    &= \frac{1}{140} \left(
        3 \big\langle I_2^2 \big\rangle
        - 12 \big\langle I_6 \big\rangle
        - 2 \big\langle I_2 I_3 \big\rangle
        + 8 \big\langle I_3^2 \big\rangle
    \right), \\
\langle A_{21}^6\rangle &= \frac{1}{24024}\Big(
    -45 \langle I_2^3 \rangle
    + 528 \langle I_2^2 I_3 \rangle
    + 208 \langle I_2 I_6 \rangle
    - 200 \langle I_2 I_3^2 \rangle \nonumber \\
&\quad 
    + 184 \langle I_4^2 \rangle
    + 64 \langle I_4 I_5 \rangle
    - 320 \langle I_5^2 \rangle
    - 2080 \langle I_6 I_3 \rangle
    + 800 \langle I_3^3 \rangle
\Big).
\end{align}

The expressions for the second moment agrees with well-known result \citep{Pope}, and the fourth-order moment is consistent with previously published findings \citep{siggia1981invariants, hierro2003fourth}.

\subsubsection{Compressible turbulence}
In this subsection, we present the expressions computed from our symbolic code for $\langle A_{21}^n\rangle$ with $n=2,4,6$ in compressible turbulence:
\begin{align}
\big\langle A_{21}^2 \big\rangle
    &= \frac{2}{15}\langle I_3\rangle - \frac{1}{15} \langle I_1^2 \rangle,\\
\langle A_{21}^4\rangle &= \frac{1}{420} \Big(
    7 \langle I_1^4 \rangle 
    + 9 \langle I_2^2 \rangle
    + 48 \langle I_1 I_5 \rangle
    - 36 \langle I_6 \rangle 
    - 6 \langle I_2 I_3 \rangle 
    + 24 \langle I_3^2 \rangle \nonumber\\
&\qquad\qquad 
    - 12 \langle I_1^2 I_2 \rangle
    - 30 \langle I_1^2 I_3 \rangle
\Big), \\
\langle A_{21}^6\rangle &= \frac{1}{72072}\Big(
    -339\langle I_1^6\rangle
    + 288\langle I_1^3 I_4\rangle
    - 2880\langle I_1^3 I_5\rangle
    + 427\langle I_1^4 I_2\rangle
    + 2624\langle I_1^4 I_3\rangle \nonumber \\
&\quad 
    -1168\langle I_1 I_2 I_4\rangle
    -384\langle I_1 I_2 I_5\rangle
    -128\langle I_1 I_4 I_3\rangle
    +9600\langle I_1 I_5 I_3\rangle +255\langle I_1^2 I_2^2\rangle \nonumber\\
&\quad +1872\langle I_1^2 I_6\rangle
    -2064\langle I_1^2 I_2 I_3\rangle
    -5400\langle I_1^2 I_3^2\rangle     -135\langle I_2^3\rangle
    +552\langle I_4^2\rangle +192\langle I_4 I_5\rangle \nonumber \\
&\quad
    -960\langle I_5^2\rangle +624\langle I_2 I_6\rangle
    +1584\langle I_2^2 I_3\rangle
    -6240\langle I_6 I_3\rangle -600\langle I_2 I_3^2\rangle
    +2400\langle I_3^3\rangle
\Big).
\end{align}

The expressions for the second moment is consistent with the result reported in \cite{yang2022low}.
        
\section{Comparison to results from numerical simulations}\label{Sec:numerical}
To test our analytical formulas, we perform numerical simulations of both incompressible and compressible turbulence.

To compute the $n$th-order moments of longitudinal and transverse velocity gradients, we proceed as follows. For each instantaneous velocity field, we first compute the spatial average of the $n$th power of each component of the velocity gradient tensor $A_{ij}$. The longitudinal moments, $\langle A_{\parallel}^n \rangle$, are defined as the average of the three diagonal components: $\langle A_{11}^n \rangle$, $\langle A_{22}^n \rangle$, and $\langle A_{33}^n \rangle$. The transverse moments, $\langle A_{\perp}^n \rangle$, are obtained by averaging the six off-diagonal components, $\langle A_{ij}^n \rangle$ for $i \ne j$. These component-wise statistics are then averaged over multiple time steps spanning at least one integral timescale to obtain the final values of $\langle A_{\parallel}^n \rangle$ and $\langle A_{\perp}^n \rangle$.
Similarly, the ensemble average of invariant quantities is computed by first evaluating their spatial average in each instantaneous field, followed by averaging over time.

For the longitudinal velocity gradient, all moments are normalized using the same reference quantity: the second-order moment $\langle A_{\parallel}^2 \rangle_{\mathrm{simu}}$ computed from the data. The standardized moments are defined as
\[
M^{\parallel}_{\mathrm{simu}} = \frac{\langle A_{\parallel}^n \rangle_{\mathrm{simu}}}{\langle A_{\parallel}^2 \rangle_{\mathrm{simu}}^{n/2}}, \quad
M^{\parallel}_{\mathrm{inv}} = \frac{\langle A_{\parallel}^n \rangle_{\mathrm{inv}}}{\langle A_{\parallel}^2 \rangle_{\mathrm{simu}}^{n/2}},
\]
where $\langle A_{\parallel}^n \rangle_{\mathrm{simu}}$ denotes the $n$th-order moment computed directly from the datasets, and $\langle A_{\parallel}^n \rangle_{\mathrm{inv}}$ is obtained from the invariant-based expressions introduced in Section~\ref{Sec:moments_A11}.
The relative error between the direct results and the invariant-based predictions is defined as
\[
\Delta_{\parallel} = \left| \frac{ M^{\parallel}_{\mathrm{simu}} - M^{\parallel}_{\mathrm{inv}} }{M^{\parallel}_{\mathrm{simu}}} \right|.
\]
Similarly, the standardized moments for transverse velocity gradients, obtained directly from the data and from invariant-based expressions, are defined as
\[
M^{\perp}_{\mathrm{simu}} = \frac{\langle A_{\perp}^n \rangle_{\mathrm{simu}}}{\langle A_{\perp}^2 \rangle_{\mathrm{simu}}^{n/2}}, \quad
M^{\perp}_{\mathrm{inv}} = \frac{\langle A_{\perp}^n \rangle_{\mathrm{inv}}}{\langle A_{\perp}^2 \rangle_{\mathrm{simu}}^{n/2}},
\]
where the theoretical moments $\langle A_{\perp}^n \rangle_{\mathrm{inv}}$ are derived from the invariant-based formulation introduced in Section~\ref{Sec:moments_A21}. The relative error is defined as
\[
\Delta_{\perp} = \left| \frac{ M^{\perp}_{\mathrm{simu}} - M^{\perp}_{\mathrm{inv}} }{M^{\perp}_{\mathrm{simu}}} \right|.
\]

\subsection{Incompressible turbulence}

\subsubsection{Numerical setups}
We perform post-processing on datasets generated with TurTLE and previously published in \cite{bentkamp2019persistent}. Direct numerical simulation (DNS) datasets are used, with Taylor-scale Reynolds numbers of $R_{\lambda} \approx 350$. The simulations were carried out on grids of size $2048^3$. The resolution is characterized by the product $k_{\mathrm{max}}\eta=3.0$, where $k_{\mathrm{max}}$ is the maximum wave number and $\eta$ is the Kolmogorov length scale. To mitigate potential bias from relying on a single dataset, we additionally analyse two independent datasets, and the corresponding results are presented in Appendix~\ref{Sec:2other_cases}.

The turbulence fields are incompressible, homogeneous, and isotropic, corresponding to fully developed, statistically stationary states. Large-scale forcing is applied to sustain stationarity by balancing the mean energy input and dissipation rates.

\subsubsection{Numerical results}
\begin{table}
  \begin{center}
\def~{\hphantom{0}}
\begin{tabular}{lcccc}

n                                                     &$M^{\parallel}_{\mathrm{simu}}$& $M^{\parallel}_{\mathrm{inv}}$& $\Delta_{\parallel}$& Contribution      \\ \hline
2                      
                                                         & 59.201          & 59.289          & 0.15\%          &                   \\ 
                                          \hline
3 
                                                          & -0.57913          & -0.58266          & 0.610\%          &                   \\ 
                                         \hline
4                  
                                     & 7.7683            & 7.7904            & 0.285\%          &                   \\ 
                                         \hline
5               
                          & -19.723           & -19.806           & 0.42\%          &                   \\ 
                                         \hline
6                    
                                       & 276.85 & 278.27 & 0.513\% & $\frac{40}{3003} \langle\operatorname{tr}(\mathsfbi{S}^2)^3\rangle$: 92.0\%; $\frac{128}{9009} \langle\operatorname{tr}(\mathsfbi{S}^3)^2\rangle$: 8.0\%   
                                         \\
                                         \hline
7                 
                                                 & -1885.6           & -1906.1           &1.09\%            &                   \\ 
                                          \hline
8                   
                                         & 37194 & 38022 & 2.23\% &$\frac{112}{21879}\langle\operatorname{tr}(\mathsfbi{S}^2)^4\rangle$: 80.6\%; $\frac{1024}{65637}\langle\operatorname{tr}(\mathsfbi{S}^2)\operatorname{tr}(\mathsfbi{S}^3)^2\rangle$: 19.4\%
                                         \\ \hline

\end{tabular}
\caption{Standardized $n$th-order moments of the longitudinal velocity gradient for the incompressible DNS results. 
Only the second-order moments are reported in unnormalized form. 
Also shown are the contributions of individual invariant terms to the sixth- and eighth-order moments.}
\label{Table:A11}
\end{center}
\end{table}

Table~\ref{Table:A11} presents the standardized longitudinal moments for various orders $n$. For $n=2$, the unnormalized values are reported. For all orders, the relative error $\Delta_{\parallel}$ remains below 2.3\%, indicating excellent agreement between the directly computed results and the invariant-based predictions, up to $n = 8$.

In addition, Table~\ref{Table:A11} shows the contributions of individual invariant terms to the sixth- and eighth-order longitudinal moments. Among all terms, those composed solely of powers of $\mathrm{tr}(\mathsfbi{S}^2)$ exhibit the largest contributions. For instance, the term $\langle \mathrm{tr}(\mathsfbi{S}^2)^3 \rangle$ contributes 92\% to the sixth-order moment, while $\langle \mathrm{tr}(\mathsfbi{S}^2)^4 \rangle$ contributes 80.6\% to the eighth-order moment. Although the contributions from $\mathrm{tr}(\mathsfbi{S}^3)$ terms are smaller, they remain non-negligible and play a noticeable role in the total moment values. Furthermore, the relative error for each order is at least one order of magnitude smaller than the contribution of any individual invariant term of that order, thereby providing strong numerical validation of the analytical formulas.

\begin{table}
  \begin{center}
\def~{\hphantom{0}}
\begin{tabular}{lcccc}

n                               & $M^{\perp}_{\mathrm{simu}}$  & $M^{\perp}_{\mathrm{inv}}$  & $\Delta_{\perp}$             & Contribution      \\ \hline
2 
                                     & 118.62         &118.58          &0.03\%           &                   \\ 
                    \hline
\multirow{4}{*}{4}  
                     & \multirow{4}{*}{11.964} & \multirow{4}{*}{11.968} & \multirow{4}{*}{0.03\%} &     $\frac{3}{140} \langle I_2^2 \rangle$: 10.99\%  \\ 
                                        &                   &                   &                   & $-\frac{3}{35}\langle I_6 \rangle$: $-30.36\%$ \\ 
                                        &                   &                   &                   &$- \frac{1}{70}\langle I_2 I_3\rangle$: 5.71\%\\ 
                                        &                   &                   &                   &$\frac{2}{35}\langle I_3^2 \rangle$: 113.66\%\\ 
                     \hline
\multirow{5}{*}{6} 
                     & \multirow{4}{*}{797.11} & \multirow{4}{*}{801.05} & \multirow{4}{*}{0.494\%} &     $-\frac{15}{8008} \langle I_2^3 \rangle$: 1.43\%; $\frac{100}{3003} \langle I_3^3 \rangle$: 128.95\%\\  
                                        &                   &                   &                   &$\frac{23}{3003} \langle I_4^2\rangle$: 0.20\%;  $ - \frac{40}{3003} \langle I_5^2 \rangle$: $-0.57\%$\\ 
                                        &                   &                   &                   &$\frac{2}{231} \langle I_2 I_6\rangle$: $-3.68\%$;  $\frac{8}{3003} \langle I_4 I_5 \rangle$: $-0.01\%$\\  &                   &                   &                   &$- \frac{20}{231} \langle I_3 I_6 \rangle$: $-64.39\%$; $\frac{2}{91} \langle I_2^2 I_3 \rangle$: 26.15\%\\ 
                                        &                   &                   &                   &$ - \frac{25}{3003} \langle I_2 I_3^2\rangle$: 11.92\%\\  
                    
                     \hline

\end{tabular}
\caption{Standardized even-order transverse velocity-gradient moments for the incompressible DNS results (second order unnormalized), together with the contributions of individual invariant terms to the fourth- and sixth-order moments.}
\label{Table:A21}
\end{center}
\end{table}

Table~\ref{Table:A21} presents the standardized transverse moments for various even orders $n$. For $n=2$, the unnormalized second-order moments are also given. For all reported orders (up to n=6), the relative error $\Delta_{\perp}$ remains below $0.5\%$, demonstrating excellent agreement with theoretical predictions. The expression for the 8th-order moment is omitted due to its prohibitive complexity.

Furthermore, Table~\ref{Table:A21} shows the contributions of individual invariant terms to the fourth- and sixth-order transverse moments. Among these, the dominant contributions arise from powers of $I_3$. Specifically, $\langle I_3^2 \rangle$ accounts for 113.66\% of the fourth-order moment, and $\langle I_3^3 \rangle$ contributes 128.95\% to the sixth-order moment, exceeding 100\% due to compensating negative contributions from other terms. The second most significant contributions, though negative, come from $\langle I_6 \rangle$ ($-30.36\%$) for $n=4$ and $\langle I_3 I_6 \rangle$ ($-64.39\%$) for $n=6$. In contrast, the contributions from $I_4$ and $I_5$ are negligible: the absolute values of $\langle I_4^2 \rangle$, $\langle I_5^2 \rangle$, and $\langle I_4 I_5 \rangle$ are all below 0.6\% for the sixth-order moment. Moreover, the relative error is at least one order of magnitude smaller than the contribution of any individual invariant term, with the exception of the three weakest terms $\langle I_4^2 \rangle$, $\langle I_5^2 \rangle$, and $\langle I_4 I_5 \rangle$, whose contributions are comparable to or smaller than the numerical uncertainty. Overall we interpret the numerical results as solid support for the analytical formulation.

\subsection{Compressible turbulence}
\subsubsection{Numerical setups}

For compressible turbulence, we solve the compressible Navier-Stokes equations numerically with the high-order finite difference solver ASTR, previously tested in DNS of various compressible turbulent flows with and without shocks \citep{fang2015direct,fang2020turbulence,yang2022low,luo2025contribution,fang2025spatially}. The details of this solver and the solved non-dimensional governing equations are presented in \cite{luo2025contribution}. Especially, for the artificial forcing term, we apply to the momentum equation an artificial Taylor-Green forcing $\rho f_i$, with 
\begin{align}
    f_1 =& A_m \cos(k_0 x_1) \sin(2k_0 x_2) \sin(-3k_0 x_3)+ A_m \cos(2k_0 x_1) \sin(-3k_0 x_2) \sin(k_0 x_3) \nonumber\\ 
    &+ A_m \cos(-3k_0 x_1) \sin(k_0 x_2) \sin(2k_0 x_3), \\
    f_2 =& A_m \sin(k_0 x_1) \cos(2k_0 x_2) \sin(-3k_0 x_3)+ A_m \sin(2k_0 x_1) \cos(-3k_0 x_2) \sin(k_0 x_3) \nonumber\\ 
    &+ A_m \sin(-3k_0 x_1) \cos(k_0 x_2) \sin(2k_0 x_3), \\
    f_3 =& A_m \sin(k_0 x_1) \sin(2k_0 x_2) \cos(-3k_0 x_3)+ A_m \sin(2k_0 x_1) \sin(-3k_0 x_2) \cos(k_0 x_3) \nonumber\\ 
    &+ A_m \sin(-3k_0 x_1) \sin(k_0 x_2) \cos(2k_0 x_3),
\end{align}
where $A_m$ is the forcing amplitude and $k_0=1$ the forcing base wave number. The forcing is thus concentrated on large scales, with wavenumbers $k\leq 3 k_0$. This forcing allows for the generation of statistically steady, homogeneous and quasi-isotropic compressible turbulence \citep{fang2025spatially}. We also add an artificial term $-\kappa T$ in the energy equation to remove the energy injected by the forcing and stabilise the computation, with the radiation coefficient calculated as $\kappa = \langle \rho f_i u_i\rangle / \langle T\rangle$. 

We perform one case in a $(2\pi)^3$ cube computational domain, discretised
with a $512^3$ mesh. Periodic boundary conditions are applied in all three directions. After entering the fully developed, statistically stationary stage, the Taylor-scale Reynolds number is $R_\lambda \approx 112$. 
The Mach number 
is $M_t \approx 0.35$. At this Mach number, local shocklets are observed within the turbulent flow field. The Kolmogorov length scale $\eta$ verifies $\eta/\Delta x = 0.52\geq 0.5$, with $\Delta x$ the grid spacing, ensuring full resolution of flow structures \citep{petersen2010forcing}.

\subsubsection{Numerical results}

\begin{table}
\begin{center}
\def~{\hphantom{0}}
\begin{tabular}{lcccc}
n              &$M^{\parallel}_{\mathrm{simu}}$& $M^{\parallel}_{\mathrm{inv}}$& $\Delta_{\parallel}$& Main contributions ($\geq 0.01\%$)      \\ \hline
2                  & 24.367                 & 23.786                 & 2.38\%                  & $\frac{2}{15} \langle\operatorname{tr}(\mathsfbi{S}^2)\rangle$: 99.96\%; $\frac{1}{15} \langle\operatorname{tr}(\mathsfbi{S})^2\rangle$: 0.04\%      \\ \hline
3                  & $-0.53125$                 & $-0.49141$                 & 7.499\%                  & $\frac{8}{105}\, \langle \mathrm{tr}(\mathsfbi{S}^3) \rangle$:100.91\%; $\frac{2}{35}\, \langle\mathrm{tr}(\mathsfbi{S}) \,\mathrm{tr}(\mathsfbi{S}^2)\rangle$:-0.91\%    \\ \hline
\multirow{2}{*}{4} & \multirow{2}{*}{5.4932}  & \multirow{2}{*}{5.1906}  & \multirow{2}{*}{5.509\%} & $\frac{4}{105} \langle \mathrm{tr}(\mathsfbi{S}^2)^2\rangle$: 100.33\%; $\frac{32}{315}\langle \mathrm{tr}(\mathsfbi{S}) \mathrm{tr}(\mathsfbi{S}^3)\rangle$: $-0.27\%$     \\ 
                   &                          &                          &                           & $-\frac{4}{105}\langle \mathrm{tr}(\mathsfbi{S})^2 \mathrm{tr}(\mathsfbi{S}^2)\rangle$:$-0.06\%$      \\ \hline
\multirow{2}{*}{5} & \multirow{2}{*}{$-10.285$} & \multirow{2}{*}{$-9.4202$} & \multirow{2}{*}{8.411\%} & $\frac{32}{693}\left\langle \mathrm{tr}(\mathsfbi{S}^2)\,\mathrm{tr}(\mathsfbi{S}^3) \right\rangle$: 100.67\%;$\frac{4}{231}\left\langle \mathrm{tr}(\mathsfbi{S})\,\mathrm{tr}(\mathsfbi{S}^2)^2 \right\rangle$: $-0.76\%$      \\
                   &                          &                          &                           & $\frac{16}{231}\left\langle \mathrm{tr}(\mathsfbi{S})^2\,\mathrm{tr}(\mathsfbi{S}^3) \right\rangle$: 0.08\%       \\ \hline
\multirow{2}{*}{6} & \multirow{2}{*}{90.37}   & \multirow{2}{*}{83.823}  & \multirow{2}{*}{7.245\%} & $\frac{40}{3003}\left\langle \mathrm{tr}(\mathsfbi{S}^2)^3 \right\rangle$: 92.84\%; $\frac{128}{9009}\left\langle \mathrm{tr}(\mathsfbi{S}^3)^2 \right\rangle$: 7.93\%     \\
                   &                          &                          &                           & $\frac{64}{1001}\left\langle \mathrm{tr}(\mathsfbi{S})\,\mathrm{tr}(\mathsfbi{S}^2)\,\mathrm{tr}(\mathsfbi{S}^3) \right\rangle$: $-0.63\%$; $-\frac{4}{143}\left\langle \mathrm{tr}(\mathsfbi{S})^2 \,\mathrm{tr}(\mathsfbi{S}^2)^2 \right\rangle$: $-0.13\% $  \\ \hline
\multirow{3}{*}{7} & \multirow{3}{*}{$-356.14$} & \multirow{3}{*}{$-331.38$} & \multirow{3}{*}{6.950\%} & $\frac{32}{1287}\left\langle \mathrm{tr}(\mathsfbi{S}^2)^2 \, \mathrm{tr}(\mathsfbi{S}^3) \right\rangle$:100.71\%; $\frac{8}{1287}\left\langle \mathrm{tr}(\mathsfbi{S}) \, \mathrm{tr}(\mathsfbi{S}^2)^3 \right\rangle$: $-0.64\%$       \\
                   &                          &                          &                           & $\frac{128}{3861}\left\langle \mathrm{tr}(\mathsfbi{S}) \, \mathrm{tr}(\mathsfbi{S}^3)^2 \right\rangle $ : $-0.16\%$; $\frac{32}{1287}\left\langle \mathrm{tr}(\mathsfbi{S})^2 \, \mathrm{tr}(\mathsfbi{S}^2)\, \mathrm{tr}(\mathsfbi{S}^3) \right\rangle $: 0.08\%      \\
                   &                          &                          &                           & $- \frac{4}{117}\left\langle \mathrm{tr}(\mathsfbi{S})^3 \, \mathrm{tr}(\mathsfbi{S}^2)^2 \right\rangle$: 0.01\%     \\ \hline
\multirow{4}{*}{8} & \multirow{4}{*}{3615.0}    & \multirow{4}{*}{3461.1}  & \multirow{4}{*}{4.259\%} & $\frac{112}{21879}\left\langle \mathrm{tr}(\mathsfbi{S}^2)^4 \right\rangle$:81.15\%; $\frac{1024}{65637}\left\langle \mathrm{tr}(\mathsfbi{S}^2)\, \mathrm{tr}(\mathsfbi{S}^3)^2 \right\rangle$:19.61\%   \\
                   &                          &                          &                           & $\frac{256}{7293}\left\langle \mathrm{tr}(\mathsfbi{S}) \, \mathrm{tr}(\mathsfbi{S}^2)^2 \, \mathrm{tr}(\mathsfbi{S}^3) \right\rangle $: $-0.50\%$; $- \frac{32}{1989}\left\langle \mathrm{tr}(\mathsfbi{S})^2 \, \mathrm{tr}(\mathsfbi{S}^2)^3 \right\rangle $: $-0.33\%$ \\
                   &                          &                          &                           & $\frac{2560}{65637}\left\langle \mathrm{tr}(\mathsfbi{S})^2 \, \mathrm{tr}(\mathsfbi{S}^3)^2 \right\rangle $: 0.08\%; $- \frac{24}{2431}\left\langle \mathrm{tr}(\mathsfbi{S})^4 \, \mathrm{tr}(\mathsfbi{S}^2)^2 \right\rangle $: $-0.01\%$                             \\
                   &                          &                          &                           & $- \frac{1280}{65637}\left\langle \mathrm{tr}(\mathsfbi{S})^3 \, \mathrm{tr}(\mathsfbi{S}^2)\, \mathrm{tr}(\mathsfbi{S}^3) \right\rangle  $: $-0.01\%$   \\ \hline                                    
\end{tabular}
\caption{Standardized $n$th-order moments of the longitudinal velocity gradient for the compressible DNS result. The second-order moments are shown in unnormalized form. Contributions from individual invariant terms to the moments at each order are also displayed; terms contributing less than $0.01\%$ are omitted.}
\label{Table:A11_com}
\end{center}
\end{table}

Table \ref{Table:A11_com} presents the standardized longitudinal moments for the compressible case for various orders $n$, with $n=2$ reporting the unnormalized values. The relative error $\Delta_{\parallel}$ is larger than that of the incompressible case, presumably mainly due to the emergence of strong local shocklets, which cause local anisotropy in the flow field. However, the relative error for all orders remains below $8.5\%$, which we consider as support of the invariant-based formulas. 

Additionally, Table \ref{Table:A11_com} also presents the contribution of individual invariant terms to each longitudinal moment. The terms with a contribution smaller than $0.01\%$ are not shown. The contribution of terms containing $\mathrm{tr}(\mathsfbi{S})$ is non-zero because of the compressibility, while remains small due to a relatively low turbulent Mach number. However, they remain non-negligible and show the difference between compressible and incompressible turbulence. Terms with $\mathrm{tr}(\mathsfbi{S}^2)$ still dominates the contribution. The ratio between $\frac{40}{3003}\left\langle \mathrm{tr}(\mathsfbi{S}^2)^3 \right\rangle$ and $\frac{128}{9009}\left\langle \mathrm{tr}(\mathsfbi{S}^3)^2 \right\rangle$ for the sixth-order is slightly larger than $23:2$ of the incompressible case, and the ratio between  $\frac{112}{21879}\left\langle \mathrm{tr}(\mathsfbi{S}^2)^4 \right\rangle$ and $\frac{1024}{65637}\left\langle \mathrm{tr}(\mathsfbi{S}^2)\, \mathrm{tr}(\mathsfbi{S}^3)^2 \right\rangle$ for the eighth-order is slightly larger than $4:1$ of the incompressible case. 

\begin{table}
  \begin{center}
\def~{\hphantom{0}}
\begin{tabular}{lcccc}
n                 & $M^{\perp}_{\mathrm{simu}}$  & $M^{\perp}_{\mathrm{inv}}$  & $\Delta_{\perp}$             & Main contributions ($\geq 0.01\%$)     \\ \hline
2                  & 47.236                & 47.527                 & 0.616\%                  & $\frac{2}{15}\langle I_3\rangle$:100.02\%; $- \frac{1}{15} \langle I_1^2 \rangle$:$-0.02\%$                   \\  \hline 
\multirow{2}{*}{4} & \multirow{2}{*}{7.8365} & \multirow{2}{*}{7.9564} & \multirow{2}{*}{1.530\%} & $\frac{2}{35}\langle I_3^2 \rangle $: 113.02\%; $-\frac{3}{35}\langle I_6 \rangle $: $-25.88\%$ ; $\frac{3}{140}\langle I_2^2 \rangle $: 8.94\% \\
                   &                         &                         &                           & $-\frac{1}{70}\langle I_2 I_3 \rangle $: 4.04\% ;$\frac{4}{35}\langle I_1 I_5 \rangle $: $-0.08\%$; $-\frac{1}{14}\langle I_1^2 I_3 \rangle $: $-0.04\%$         \\\hline   
\multirow{5}{*}{6} & \multirow{5}{*}{222.31} & \multirow{5}{*}{233.23} & \multirow{5}{*}{4.915\%} & $\frac{100}{3003}\langle I_3^3\rangle $: 121.60\%; $-\frac{20}{231}\langle I_6 I_3\rangle $: $-44.67\%$ ; $\frac{2}{91}\langle I_2^2 I_3\rangle $: 17.07\%        \\
                   &                         &                         &                           & $\frac{25}{3003}\langle I_2 I_3^2\rangle $: 7.90\% ; $\frac{2}{231}\langle I_2 I_6\rangle $: $-1.89\%$; $-\frac{15}{8008}\langle I_2^3\rangle $: 0.72\% \\
                   &                         &                         &                           & $-\frac{40}{3003}\langle I_5^2\rangle $: $-0.64\%$; $\frac{400}{3003}\langle I_1 I_5 I_3\rangle$: $-0.19\%$ ;  $\frac{23}{3003}\langle I_4^2\rangle $: 0.17\%   \\
                   &                         &                         &                           & $-\frac{75}{1001}\langle I_1^2 I_3^2\rangle $: $-0.08\%$; $-\frac{146}{9009}\langle I_1 I_2 I_4\rangle$: 0.01\%; $-\frac{86}{3003}\langle I_1^2 I_2 I_3\rangle $: 0.01\% \\
                   &                         &                         &                           &  $\frac{8}{3003}\langle I_4 I_5\rangle $: $-0.01\%$ \\
                   \hline
\end{tabular}
\caption{Standardized even-order transverse velocity-gradient moments for the compressible DNS result (second order unnormalized), together with the contributions of individual invariant terms to each-order moments; terms contributing less than $0.01\%$ are omitted.}
\label{Table:A21_com}
\end{center}
\end{table}

Table \ref{Table:A21_com} presents the standardized transverse moments for the compressible case, where the unnormalized second-order moments are shown for $n=2$. Similarly to the longitudinal moments, the relative error $\Delta_{\parallel}$ is larger than in the incompressible case, while it remains below $5.0\%$ and indicates good agreement with theoretical predictions up to $n=6$. 

Moreover, Table \ref{Table:A21_com} also reports the contributions of individual invariant terms to each transverse moment, with the contributions smaller than $0.01\%$ omitted. Terms involving $I_1$ appear in the compressible case, although they remain small due to the low turbulent Mach number. Contributions associated with powers of $I_3$ continue to dominate the fourth- and sixth-order transverse moments: $\langle I_3^2 \rangle$ accounts for $113.02\%$ of the fourth-order moment, while  $\langle I_3^3 \rangle$  contributes $121.60\%$ to the sixth-order moment. These proportions are of the same order of magnitude as those observed in the incompressible case. The second and third most significant contributions arise from $\langle I_6 \rangle (-25.88\%)$  for $n=4$ and $\langle I_3 I_6 \rangle(-44.67\%)$ for $n=6$. Although these contributions remain negative, their magnitudes are significantly smaller than in incompressible turbulence. A similar reduction is observed for other terms, such as $\langle I_2^2 \rangle$ for $n=4$ and $\langle I_2^2 I_3 \rangle$ for $n=6$, indicating a weakened influence of these higher-order interactions in the compressible regime.

\section{Conclusion}\label{Sec:conclu}
In this work, we investigated the statistics of velocity gradients from the perspective of velocity gradient invariants. We developed an analytical framework that expresses the \(n\)th-order moments of longitudinal and transverse velocity gradients in terms of statistical velocity gradient invariants, without the need to solve complex systems of equations as required in traditional approaches. This offers a more direct way to analyze high-order statistics in turbulence.

The analytical results were validated by numerical simulations for incompressible and compressible turbulence, comparing statistical moments of longitudinal (up to 8th order) and transverse (up to 6th order) velocity gradients. For incompressible turbulence, the relative errors for longitudinal moments are generally below \(5\%\), while those for transverse moments are even smaller—typically below \(0.5\%\). For compressible turbulence, despite the slight anisotropy induced by local shocklets, the relative errors for longitudinal moments remain below $8.5\%$, and those for transverse moments remain below $5\%$. This agreement supports the validity of our formulation within the tested range.

To the best of our knowledge, explicit analytical expressions for velocity gradient moments beyond the fourth order have only recently become available, restricted to the longitudinal incompressible case \citep{yang2026relation}. In this work, we present an independent derivation based on a different approach, yielding explicit expressions for both longitudinal and transverse velocity gradients in compressible and incompressible turbulence. In particular, we obtain a closed-form expression for longitudinal moments valid for arbitrary order (Eq.~\eqref{Eq:A_11_n}).

We also quantified the contributions of individual invariant terms to the high-order velocity gradient moments. 
For longitudinal velocity gradients, the invariant terms composed solely of powers of $\mathrm{tr}(\mathsfbi{S}^2)$ exhibit the largest contributions. However, the terms involving $\mathrm{tr}(\mathsfbi{S}^3)$ also contribute noticeably and cannot be neglected.
For transverse velocity gradients, the dominant contributions come from invariant terms composed solely of powers of $\mathrm{tr}(\mathsfbi{A} \mathsfbi{A}^T)$. The second-largest contributions arise from terms involving $\mathrm{tr}(\mathsfbi{A} \mathsfbi{A} \mathsfbi{A}^T \mathsfbi{A}^T)$, although these are negative. In contrast, invariant terms involving $\mathrm{tr}(\mathsfbi{A}^3)$ and $\mathrm{tr}(\mathsfbi{A} \mathsfbi{A} \mathsfbi{A}^T)$ are negligible.

Overall, our results provide a theoretical foundation for analyzing high-order velocity gradient statistics using invariant-based formulations. These exact expressions may serve as physical constraints in the development of velocity gradient models—any such model must remain consistent with these formal relations. They may also be employed to assess the degree of isotropy in turbulent flows by testing the consistency of measured high-order moments with the derived invariant structures. As computational resources continue to grow, enabling increasingly accurate evaluations of high-order moments in numerical simulations, having closed-form analytical expressions becomes particularly valuable for interpreting and validating such data.

For future work, we anticipate that our approach can be extended to compute arbitrary-order moments of any combination of velocity gradient components, offering a general framework for analyzing complex turbulence statistics.

\section*{Acknowledgment}

During the preparation of this manuscript, we became aware of related and independent work by \cite{yang2026relation}, who kindly shared their results with us. The authors thank Cristian Lalescu for providing the incompressible DNS dataset, and Ke Li for the inputs.

\section*{Funding}

This project has received funding from the European Research Council (ERC) under the European Union’s Horizon 2020 research and innovation programme (Grant Agreement No.~101001081). The authors gratefully acknowledge the scientific support and high-performance computing (HPC) resources provided by the Erlangen National High Performance Computing Center (NHR@FAU) at Friedrich-Alexander-Universität Erlangen-Nürnberg (FAU) under the NHR project “EnSimTurb”. NHR funding is provided by federal and Bavarian state authorities. The NHR@FAU hardware is partially supported by the German Research Foundation (DFG) No.~440719683.

\section*{AI transparency statment}

Generative AI (ChatGPT, OpenAI) was used for language editing after a complete first draft of the manuscript had been written by the authors, and to assist in drafting initial code implementations based on author-specified algorithms. All code was subsequently modified, debugged, and validated by the authors.

\begin{appen}
\section{Formulas in terms of invariants based on $\mathsfbi{S}$ and $\mathsfbi{W}$}\label{Sec:Invariants_S_W}
In turbulence studies, invariants constructed from the strain-rate tensor 
$\mathsfbi{S}$ and the rotation-rate tensor $\mathsfbi{W}$ are widely used. 
For completeness, we present the expressions of the transverse velocity-gradient 
moments in this invariant basis.

The tensors $\mathsfbi{S}$ and $\mathsfbi{W}$ are defined as the symmetric and 
antisymmetric parts of the velocity-gradient tensor $\mathsfbi{A}$:
\[
\mathsfbi{S}= \frac{1}{2} (\mathsfbi{A} + \mathsfbi{A}^T), \quad \mathsfbi{W} = \frac{1}{2} (\mathsfbi{A} - \mathsfbi{A}^T).
\]
The invariants $(I_1',\ldots,I_6')$ are defined in Eq.~\eqref{Eq:invariants_WS}. This basis is equivalent to the $\operatorname{tr}(\mathsfbi{A})$-based invariants via
\begin{equation}
\begin{aligned}\label{Eq:relation_A_SOmega}
     \operatorname{tr}(\mathsfbi{A}) &= \operatorname{tr}(\mathsfbi{S}), \\
     \operatorname{tr}(\mathsfbi{A}^2) &= \operatorname{tr}(\mathsfbi{S}^2) + \operatorname{tr}(\mathsfbi{W}^2), \\
     \operatorname{tr}(\mathsfbi{A} \mathsfbi{A}^T) &= \operatorname{tr}(\mathsfbi{S}^2) - \operatorname{tr}(\mathsfbi{W}^2), \\
     \operatorname{tr}(\mathsfbi{A}^3) &= \operatorname{tr}(\mathsfbi{S}^3) + 3 \operatorname{tr}(\mathsfbi{S} \mathsfbi{W}^2), \\
     \operatorname{tr}(\mathsfbi{A}^2 \mathsfbi{A}^T) &= \operatorname{tr}(\mathsfbi{S}^3) - \operatorname{tr}(\mathsfbi{S} \mathsfbi{W}^2), \\
     \operatorname{tr}(\mathsfbi{A}^2 (\mathsfbi{A}^T)^2) &=   \operatorname{tr}(\mathsfbi{S}^4) +\operatorname{tr}(\mathsfbi{W}^4)  - 2\operatorname{tr}(\mathsfbi{S} \mathsfbi{W}\mathsfbi{S}\mathsfbi{W})\\
     & =\frac{1}{6} {I_1'}^4 - {I_1'}^2I_2'+\frac{1}{2}{I_2'}^2 + \frac{4}{3}I_1'I_4'+\frac{1}{2}{I_3'}^2 +I_3'(I_1'^2-I_2')+4I_6'-4I_1'I_5'.
\end{aligned}
\end{equation}
Using these relations, the transverse velocity-gradient moments can be 
expressed in terms of the invariants $(I_1',\ldots,I_6')$.
For incompressible turbulence, the resulting expressions are:
\begin{align}
    \big\langle A_{21}^2 \big\rangle &= \frac{4}{15} \langle I_2' \rangle, \\
    \big\langle A_{21}^4 \big\rangle &= \frac{1}{20} \langle I_3'^2 \rangle + \frac{1}{70} \langle I_2' I_3' \rangle +\frac{3}{140} \langle I_2'^2 \rangle - \frac{12}{35} \langle I_6' \rangle, \\
    \big\langle A_{21}^6 \big\rangle &= -\frac{1}{56} \langle I_3'^3 \rangle -\frac{3}{56} \langle I_3'^2 I_2' \rangle +\frac{31}{616} \langle I_3'I_2'^2 \rangle +\frac{7}{1144}\langle I_2'^3 \rangle +\frac{8}{21}\langle I_3'I_6' \rangle - \frac{24}{77}\langle I_2' I_6' \rangle \nonumber \\
    &\quad - \frac{3}{1001}\langle I_4'^2 \rangle +\frac{6}{77} \langle I_4'I_5' \rangle + \frac{1}{21}\langle I_5'^2 \rangle
\end{align}
For compressible turbulence, the expressions based on $\mathsfbi{S}$ and $\mathsfbi{W}$ are:
\begin{align}
    \big\langle A_{21}^2 \big\rangle &= \frac{4}{15} \langle I_2' \rangle - \frac{1}{5}\langle I_1'^2 \rangle, \\
    \big\langle A_{21}^4 \big\rangle &= \frac{1}{20} \langle I_3'^2 \rangle - \frac{3}{70} \langle I_3'I_1'^2 \rangle + \frac{1}{420} \langle I_1'^4 \rangle+\frac{1}{70} \langle I_2' I_3' \rangle - \frac{1}{70} \langle I_1'^2 I_2'\rangle+\frac{3}{140} \langle I_2'^2 \rangle  \nonumber\\
    &\quad - \frac{12}{35} \langle I_6' \rangle + \frac{8}{35} \langle I_1'I_5'\rangle, \\
    \big\langle A_{21}^6 \big\rangle &= -\frac{\langle I_3'^3 \rangle}{56}
+ \frac{11}{168} \langle I_3'^2 I_1'^2 \rangle
+ \frac{1}{88} \langle I_3' I_1'^4\rangle
- \frac{3}{8008} \langle I_1'^{6} \rangle - \frac{3}{56} \langle I_3'^2 I_2' \rangle - \frac{13}{308} \langle I_3' I_1'^2 I_2' \rangle \nonumber\\
& + \frac{27}{8008} \langle I_1'^4 I_2' \rangle
+ \frac{31}{616} \langle I_3' I_2'^2 \rangle - \frac{73}{8008} \langle I_1'^2 I_2'^2 \rangle
+ \frac{7}{1144} \langle I_2'^3 \rangle
+ \frac{8}{21} \langle I_3' I_6' \rangle  + \frac{8}{77} \langle I_1'^2 I_6' \rangle \nonumber\\
& - \frac{24}{77} \langle I_2' I_6' \rangle
- \frac{2}{77} \langle I_3' I_1' I_4' \rangle
- \frac{4}{3003} \langle I_1'^3 I_4' \rangle
+ \frac{6}{1001} \langle I_1' I_2' I_4' \rangle - \frac{3}{1001} \langle I_4'^2 \rangle
- \frac{2}{7} \langle I_3' I_1' I_5' \rangle \nonumber\\
& - \frac{4}{77} \langle I_1'^3 I_5' \rangle
+ \frac{10}{77} \langle I_1' I_2' I_5' \rangle + \frac{6}{77} \langle I_4' I_5' \rangle
+ \frac{1}{21} \langle I_5'^2 \rangle.
\end{align}

The corresponding implementation is available in the
\href{https://cocalc.com/share/public_paths/78ee59efe8b5ac10ea4d7f68c2d337e7413afea1}{JFM notebook}.

\section{Additional simulations}\label{Sec:2other_cases}

To avoid potential bias arising from reliance on a single dataset, we additionally performed post-processing on two other datasets reported in \cite{bentkamp2019persistent} with $R_{\lambda} \approx 239$ and 509, and grids of size $1024^3$ and $2048^3$, respectively, both satisfying $k_{\mathrm{max}}\eta=1.5$. The corresponding setup parameters are summarized in Table~\ref{Table:Simulation_cases}. 

\begin{table}
  \begin{center}
\def~{\hphantom{0}}
\begin{tabular}{cccc}
Case & Grid size & $R_{\lambda}$ & $k_{\mathrm{max}}\eta$   \\ 
a & $1024^3$  & 239           & 1.5   \\ 
b & $2048^3$  & 509           & 1.5  \\ 
\end{tabular}
\caption{Simulation parameters for the two additional DNS cases used in this study.}
\label{Table:Simulation_cases}
\end{center}
\end{table}

The standardized $n$th-order moments of the longitudinal velocity gradient for the additional DNS cases are shown in  Table \ref{Table:A11_2others}, and those of the transverse velocity gradient are shown in  Table \ref{Table:A21_2others}. Similar results to those reported in Tables~\ref{Table:A11} and \ref{Table:A21} are observed.

\begin{table}
  \begin{center}
\def~{\hphantom{0}}
\begin{tabular}{lccccc}
n                                        & Cases               &$M^{\parallel}_{\mathrm{simu}}$& $M^{\parallel}_{\mathrm{inv}}$& $\Delta_{\parallel}$& Contribution      \\ \hline
\multirow{2}{*}{2}                       & a                  & 63.624          & 63.621          &  0.005\%         & \multirow{3}{*}{} \\ 
                                 
                                         & b                  & 177.12         & 177.18         &0.03\%           &                   \\ \hline
\multirow{2}{*}{3}                       & a                  & -0.56202          & -0.56200          & 0.004\%          & \multirow{3}{*}{} \\ 
                                        
                                         & b                  & -0.61170          & -0.61266          & 0.16\%          &                   \\ \hline
\multirow{2}{*}{4}                       & a                  & 7.0453            & 7.0458            &  0.007\%         & \multirow{3}{*}{} \\                       
                                         & b                  & 9.2014            & 9.2120            & 0.115\%          &                   \\ \hline
\multirow{2}{*}{5}                       & a                  & -16.154           & -16.148           & 0.04\%          & \multirow{3}{*}{} \\ 
                                         
                                         & b                  & -26.047           & -26.131           & 0.32\%          &                   \\ \hline
\multirow{4}{*}{6}                       & \multirow{2}{*}{a} & \multirow{2}{*}{196.43} & \multirow{2}{*}{196.49} & \multirow{2}{*}{0.03\%} & $\frac{40}{3003} \langle\operatorname{tr}(\mathsfbi{S}^2)^3\rangle$: 92.0\%                  \\ 
                                         &                    &                   &                   &                   &$\frac{128}{9009} \langle\operatorname{tr}(\mathsfbi{S}^3)^2\rangle$: 8.0\%                    \\ \cline{2-6} 
                                        
                                         & \multirow{2}{*}{b} & \multirow{2}{*}{436.48} & \multirow{2}{*}{439.77} & \multirow{2}{*}{0.754\%} &       $\frac{40}{3003} \langle\operatorname{tr}(\mathsfbi{S}^2)^3\rangle$: 92.2\% 
                                         \\ 
                                         &                    &                   &                   &                   &$\frac{128}{9009} \langle\operatorname{tr}(\mathsfbi{S}^3)^2\rangle$: 7.8\%  
                                         \\ \hline
\multirow{2}{*}{7}                       & a                  & -1093.5           & -1092.0           &0.14\%           & \multirow{3}{*}{} \\
                                         
                                         & b                  & -3314.9           & -3389.4           &2.25\%            &                   \\ \hline
\multirow{4}{*}{8}                       & \multirow{2}{*}{a} & \multirow{2}{*}{16602} & \multirow{2}{*}{16688} & \multirow{2}{*}{0.52\%} & $\frac{112}{21879}\langle\operatorname{tr}(\mathsfbi{S}^2)^4\rangle$: 80.0\%                  \\ 
                                         &                    &                   &                   &                   & $\frac{1024}{65637}\langle\operatorname{tr}(\mathsfbi{S}^2)\operatorname{tr}(\mathsfbi{S}^3)^2\rangle$: 20.0\%\\ \cline{2-6} 
                                        
                                         & \multirow{2}{*}{b} & \multirow{2}{*}{75521} & \multirow{2}{*}{79283} & \multirow{2}{*}{4.981\%} &          $\frac{112}{21879}\langle\operatorname{tr}(\mathsfbi{S}^2)^4\rangle$: 80.4\%
                                         \\ 
                                         &                    &                   &                   &                   &$\frac{1024}{65637}\langle\operatorname{tr}(\mathsfbi{S}^2)\operatorname{tr}(\mathsfbi{S}^3)^2\rangle$: 19.6\%\\ 
                                         \hline

\end{tabular}
\caption{Standardized $n$th-order moments of the longitudinal velocity gradient for the two DNS cases. 
Only the second-order moments are reported in unnormalized form. 
Also shown are the contributions of individual invariant terms to the sixth- and eighth-order moments.}
\label{Table:A11_2others}
\end{center}
\end{table}

\begin{table}
  \begin{center}
\def~{\hphantom{0}}
\begin{tabular}{lccccc}
n                   & Cases               & $M^{\perp}_{\mathrm{simu}}$  & $M^{\perp}_{\mathrm{inv}}$  & $\Delta_{\perp}$             & Contribution      \\ \hline
\multirow{2}{*}{2}  & a                  & 127.240         &127.241          &0.0008\%           & \multirow{3}{*}{} \\ 
                    
                    & b                  & 354.40         &354.36          &0.011\%           &                   \\ \hline
\multirow{8}{*}{4} & \multirow{4}{*}{a} & \multirow{4}{*}{10.801} & \multirow{4}{*}{10.800} & \multirow{4}{*}{0.009\%} &  $\frac{3}{140} \langle I_2^2 \rangle$: 10.95\%             \\ 
                    &                    &                   &                   &                   &$-\frac{3}{35}\langle I_6 \rangle$: $-30.35\%$\\ 
                    &                    &                   &                   &                   &$- \frac{1}{70}\langle I_2 I_3\rangle$: 5.64\%\\  
                    &                    &                   &                   &                   & $\frac{2}{35}\langle I_3^2 \rangle$: 113.76\% \\ \cline{2-6} 
                     
                    & \multirow{4}{*}{b} & \multirow{4}{*}{14.336} & \multirow{4}{*}{14.337} & \multirow{4}{*}{0.007\%} &     $\frac{3}{140} \langle I_2^2 \rangle$: 11.34\%  \\ 
                    &                    &                   &                   &                   &$-\frac{3}{35}\langle I_6 \rangle$: $-31.29\%$ \\ 
                    &                    &                   &                   &                   &$- \frac{1}{70}\langle I_2 I_3\rangle$: 5.99\%     \\ 
                    &                    &                   &                   &                   &$\frac{2}{35}\langle I_3^2 \rangle$: 113.96\% \\ \hline
\multirow{8}{*}{6} & \multirow{4}{*}{a} & \multirow{4}{*}{581.17} & \multirow{4}{*}{579.69} & \multirow{4}{*}{0.255\%} &                          $-\frac{15}{8008} \langle I_2^3 \rangle$: 1.44\%; $\frac{100}{3003} \langle I_3^3 \rangle$: 129.38\% \\ 
                    &                    &                   &                   &                   &$\frac{23}{3003} \langle I_4^2\rangle$: 0.21\%;  $ - \frac{40}{3003} \langle I_5^2 \rangle$: $-0.56\%$ \\ 
                    &                    &                   &                   &                   &$\frac{2}{231} \langle I_2 I_6\rangle$: $-3.73\%$;  $\frac{8}{3003} \langle I_4 I_5 \rangle$: $-0.01\%$ \\ 
                    &                    &                   &                   & 
                    &$- \frac{20}{231} \langle I_3 I_6 \rangle$: $-65.70\%$; $\frac{2}{91} \langle I_2^2 I_3 \rangle$: 26.71\% \\ 
                    &                    &                   &                   &                   &$ - \frac{25}{3003} \langle I_2 I_3^2\rangle$: 12.26\%\\ \cline{2-6} 
                    
                    & \multirow{4}{*}{b} & \multirow{4}{*}{1341.2} & \multirow{4}{*}{1343.2} & \multirow{4}{*}{0.15\%} &     $-\frac{15}{8008} \langle I_2^3 \rangle$: 1.62\%; $\frac{100}{3003} \langle I_3^3 \rangle$: 131.77\% \\ 
                    &                    &                   &                   &                   &$\frac{23}{3003} \langle I_4^2\rangle$: 0.22\%;  $ - \frac{40}{3003} \langle I_5^2 \rangle$: $-0.53\%$\\  
                    &                    &                   &                   &                   &$\frac{2}{231} \langle I_2 I_6\rangle$: $-4.18\%$;  $\frac{8}{3003} \langle I_4 I_5 \rangle$: $-0.01\%$ \\
                    &                    &                   &                   &                   &$- \frac{20}{231} \langle I_3 I_6 \rangle$: $-71.50\%$; $\frac{2}{91} \langle I_2^2 I_3 \rangle$: 29.23\%\\ 
                    &                    &                   &                   &                   &$ - \frac{25}{3003} \langle I_2 I_3^2\rangle$: 13.38\% \\ \hline

\end{tabular}
\caption{Standardized even-order transverse velocity-gradient moments for the two DNS cases (second order unnormalized), together with the contributions of individual invariant terms to the fourth- and sixth-order moments.}
\label{Table:A21_2others}
\end{center}
\end{table}

\end{appen}\clearpage

\bibliographystyle{jfm}
\bibliography{biblio}

\end{document}